\documentstyle[eqsecnum,pre,aps,epsfig,epsf,multicol]{revtex}

\def\pv{{\bf p}}
\def\nv{{\bf n}}
\def\mv{{\bf m}}
\def\xv{{\bf x}}
\def\lv{{\bf l}}
\def\gradv{{\bf \nabla}}
\newcommand{\remove}[1]{}

\begin{document}

\title{Theory of Banana Liquid Crystal Phases and Phase Transitions}
\author{T. C. Lubensky}
\address{Department of Physics, University of Pennsylvania,
Philadelphia, Pennsylvania 19174}
\author{Leo Radzihovsky}
\address{Department of Physics, University of Colorado,
Boulder, CO 80309}

\date{\today}
\maketitle
\begin{abstract}

  We study phases and phase transitions that can take place in the
  newly discovered banana (bow-shaped or bent-core) liquid crystal molecules. 
  We show that to completely characterize phases exhibited by such bent-core
  molecules a third-rank tensor $T^{ijk}$ order parameter is
  necessary in addition to the vector and the nematic (second-rank)
  tensor order parameters.  We present an exhaustive list of possible
  liquid phases, characterizing them by their space-symmetry group and
  order parameters, and catalog the universality classes of the
  corresponding phase transitions that we expect to take place in such
  bent-core molecular liquid crystals.  In addition to the
  conventional liquid-crystal phases such as the nematic phase, we predict
  the existence of novel liquid phases, including the
  spontaneously chiral nematic $(N_T + 2)^*$ and chiral polar $(V_T +
  2)^*$ phases, the orientationally-ordered but optically isotropic
  tetrahedratic $T$ phase, and a novel nematic $N_T$ phase with $D_{2d}$
  symmetry that is neither uniaxial nor biaxial.  Interestingly, the
  Isotropic-Tetrahedratic transition is {\em continuous} in mean-field
  theory, but is likely driven first-order by thermal fluctuations. We
  conclude with a discussion of smectic analogs of these phases and their
  experimental signatures.

\end{abstract}
\begin{multicols}{2}
\narrowtext

\section{Introduction}
\label{Introduction} Liquid crystals are extraordinary systems in
that they continue to have a revolutionary technological impact
and to consistently pose new theoretical challenges of fundamental
interest. They exhibit a rich variety of phases with symmetries
intermediate between those of a the highest symmetry homogeneous
isotropic liquid and the lowest-symmetry three-dimensional
periodic crystal. In contrast to their magnetic and ferroelectric
{\it solid state} analogs, whose ordering is driven by
energy-entropy competition, liquid-crystal phase transitions are
of predominantly entropic origin. Not unrelated to this is the
fact that, with one exception of the chiral smectic-$C^*$
phase\cite{Meyer}, commonly observed liquid crystal phases are
non-polar. It is, therefore, not surprising that a recent
experimental discovery by the Tokyo Tech group\cite{Watanabe} of
ferroelectricity in the liquid crystal phase of {\it achiral}
bend-core (banana) molecules has captured the attention of the
liquid crystal community. Subsequent light microscopy studies by
the Boulder group\cite{Link} elucidated the molecular organization
of the newly discovered phase. They convincingly demonstrated that
(what came to be known as) the $B_2$ phase\cite{B2phases} is an
antiferroelectric smectic-$C$ phase in which layers {\em
spontaneously} break chiral symmetry (with chirality alternating
from layer to layer) and exhibit polar order in achiral molecules.
Eight distinct phases of bent-core molecules, tentatively labeled
$B_1$ to $B_8$ have been identified\cite{Bphases,B8}, though most have not
been fully characterized. Two of the most well studied the $B_2$ and $B_7$,
are smectic phases consisting of stacks of fluid layers
with some internal tilt order, and are special because they can be switched
with an electric field. A material composed of {\em
achiral} nematogens having a ground state that is ferroelectric
and {\em homogeneously} chiral has also recently been
discovered.\cite{Walba} This experimental discovery opens up a
vast new class of achiral molecules that nevertheless exhibit
ferroelectricity and are therefore of interest to the liquid
crystal display technology.

While the study of banana liquid crystals has seen substantial
experimental strides\cite{Meyer,Watanabe,Link,BerlinWorkshop},
there has been relatively little basic theoretical work on this
fascinating new class of materials. Brand {\it et
al.}\cite{Brand} presented an exhaustive, model-independent
classification of the symmetry-allowed smectic phases, and Roy
{\it et al.}\cite{Roy} introduced a phenomenological Landau model
that produces many of the banana smectic phases.  There are also a
number of numerical simulations\cite{Neal,Camp,Billeter,Memmer,Glaser} on
systems of model bent-core molecules that produce nematic phases
as well as some of the possible smectic phases. Because so far,
experimental examples of orientationally-ordered but spatially
homogeneous phases {\em liquid} phases\cite{bananaNematics} are
rare, most of the efforts have focussed on the {\em smectic}
phases of bent-core molecules. Here, we will instead focus on
spatially homogeneous phases, which we will refer to as {\em
liquid} phases, the understanding and classifying of whose phase
behavior is in many ways a prerequisite to the study of more
ordered (e.g., smectic) phases, which in addition break
translational symmetry. We will take advantage of the formal
developments and analysis presented in this paper for the liquid
phases in our studies of smectic phases, which we defer to a
future publication.\cite{future}

The first primary conclusion of our work, which forms the starting
point of all further analysis presented here, is that a {\em
third}-rank traceless symmetric tensor order parameter $T^{ijk}$,
in addition to the usual nematic $Q^{ij}$ and vector $p^i$ order
parameters, is necessary in order to capture the orientational
order observed in experiments on banana molecules. Without
introducing such an angular momentum $L=3$ order parameter, only
structures which have {\em at least} $C_2$ and mirror symmetry,
such as e.g., the biaxial nematic can be captured, thereby
precluding a first-principles order parameter description of, for
example, the most interesting spontaneously-ordered {\em chiral}
phases.

As we will demonstrate in great detail, once this higher-order
order parameter $T^{ijk}$ is introduced, a complex web (displayed
in Fig.\ \ref{flowchart}) of possible liquid phases emerges and,
associated with them, a very rich phase behavior. Many of these
phases exhibit exotic symmetries summarized in Table 1, including
D$_{2d}$, D$_2$, and C$_2$ symmetries\cite{Tinkham}, which have
not to our knowledge been previously identified in spatially
uniform (i.e., liquid) states. These new anisotropic liquid phases
are distinguished by the nature of their $T^{ijk}$ ordering. The
diversity in phase diagram topologies originates from a large
number of symmetry-allowed transition sequences between many of
the phases that exhibit some nontrivial combination of the $p^i$,
$Q^{ij}$ and/or $T^{ijk}$ order parameters. Some of the novel
orientationally-ordered liquid states that we predict are the
spontaneously chiral nematic $(N_T + 2)^*$ and chiral polar $(V_T
+ 2)^*$ phases, an optically isotropic tetrahedratic $T$ phase,
and a novel nematic $N_T$ phase, with $D_{2d}$ symmetry, that is
neither uniaxial nor biaxial, but rather exhibits a 4-fold
improper ($S_4$) rotational symmetry about its nematic axis.

The paper is organized as follows. In Sec.\ \ref{Model}, we
present a model of a banana-shaped liquid-crystal molecule. By
considering mass moments of molecules with this shape, we are
naturally led to introduce the three important order parameters,
$p^i$, $Q^{ij}$ and $T^{ijk}$, that are necessary to fully
describe anisotropic liquid states into which such molecules can
macroscopically order. In Sec.\ \ref{PhasesSymmetries} we then
catalog all thermodynamically distinct liquid phases
characterizable by these three order parameters. We organize these
phases according to symmetry groups under which they are invariant
and present an exhaustive list of phase transition sequences
allowed by symmetry. We construct a Landau theory of the three
order parameters in Sec.\ \ref{LandauTheory} and analyze the
nature of the complicated web of phase transitions that it
predicts in Sec.\ \ref{PhaseTransitions}, finding full consistency
with our general group-theoretic analysis.  In Sec.\
\ref{smectic}, we briefly discuss possible new smectic phases that
could result when smectic ordering develops in the various liquid
phases we identify. We conclude with Sec.\ \ref{Conclusions} by
summarizing our results and discussing their relevance to future
studies of smectic phases and to experiments.

\section{Simple Model of Banana Liquid Crystals: Order Parameters}
\label{Model}

\begin{figure}[bth]
\centering
\setlength{\unitlength}{1mm}
\begin{picture}(150,85)(0,0)
\put(-3,-65){\begin{picture}(150,200)(0,0)
\includegraphics{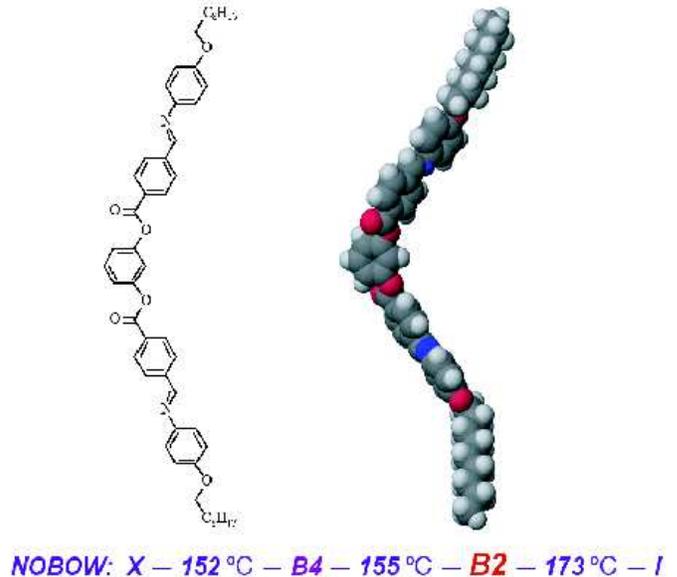}
\end{picture}}
\end{picture}
\caption{Chemically accurate model of a bent-core (banana) NOBOW molecule
  studied in Refs.\ [2] and [3] that displays Isotropic, B$_2$, B$_4$ and
  Crystal (X) phases. Instantaneously, a molecule can be found in a
  chiral nonplanar configuration (as shown on the right), but
  fluctuates equally between positive and negative chiralities, on
  average is planar and therefore achiral.}
\label{molecule}
\end{figure}

\begin{figure}[bth]
\centering
\setlength{\unitlength}{1mm}
\begin{picture}(150,60)(0,0)
\put(-20,-57){\begin{picture}(150,0)(0,0)
\includegraphics{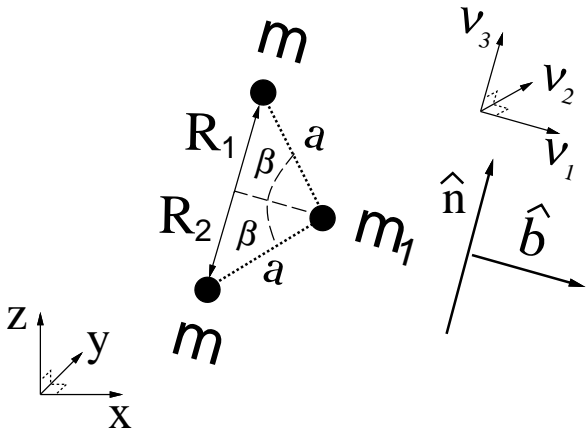}
\end{picture}}
\end{picture}
\caption{A simple three-atom model of a banana molecule
and a body-fixed orthonormal coordinate system, capturing the
molecule's characteristic, achiral $C_{2v}$ symmetry.}
\label{molecule_model}
\end{figure}

As is clear from a chemically and geometrically accurate schematic
of a bent-core molecule, shown in Fig.\ref{molecule}, the most
notable characteristic of banana molecules is their ``V'' shape
with bent, (on average) planar and therefore achiral cores. This
shape earned such molecules a name ``bow''-shaped.  The molecule
is characterized by C$_{2 v}$ symmetry, defined by a non-polar
direction ${\bbox\nu}_3$ (the ``string'' of the bow), pointing
from one endpoint of the ``V'' to the other and an orthogonal
polar axis ${\bbox\nu}_1$ (the bow's ``arrow''), pointing to the
vertex of the ``V'', as illustrated in Fig.\ \ref{molecule_model}.

We can, therefore, expect the molecule to be characterized by both
even- and odd-rank symmetric, traceless tensors with the preferred
axis of odd-rank tensors along the molecular ${\bbox\nu}_1$-axis.
We can capture these molecular features by a simple three-atom
rigid bond model of the banana molecule illustrated in Fig.\
\ref{molecule_model}.
As just discussed, associated with each banana molecule $\alpha$
is a body-fixed orthonormal coordinate system with unit vectors
(${\bbox\nu}_{\alpha,1}$, ${\bbox\nu}_{\alpha,2}$, ${\bbox
\nu}_{\alpha,3}$). Molecular C$_{2 v}$ symmetry implies invariance
under the reflection operations ${\bbox\nu}_{\alpha,3}\rightarrow
-{\bbox\nu}_{\alpha,3}$ and ${\bbox\nu}_{\alpha,2}\rightarrow -
{\bbox\nu}_{\alpha,2}$ and the $\pi$-rotation (about
${\bbox\nu}_{\alpha,1}$) operation
${\bbox\nu}_{\alpha,2}\rightarrow -{\bbox\nu}_{\alpha,2}$,
${\bbox\nu}_{\alpha,3} \rightarrow -{\bbox\nu}_{\alpha,3}$, but
not under the reflection ${\bbox\nu}_{\alpha,1}\rightarrow
-{\bbox\nu}_{\alpha,1}$.  In this body-fixed frame, the three
atoms' coordinates are given by
\begin{mathletters}
\begin{eqnarray}
{\bf R}_{\alpha,1}&=&(a\cos\beta){\bbox\nu}_{\alpha,1}\;,\\ {\bf
R}_{\alpha,2}&=&-(a\sin\beta){\bbox\nu}_{\alpha,3}\;,\\ {\bf
R}_{\alpha,3}&=&(a\sin\beta){\bbox\nu}_{\alpha,3}\;,
\label{rs1}
\end{eqnarray}
\end{mathletters}
where $2 \beta\approx 120^o$ is the opening angle of the ``V''\cite{Glaser}, 
as shown in Fig.\ \ref{molecule_model}, and with the origin located
on the ${\bbox\nu}_{\alpha,3}$ axis, half-way between $m$ masses.
As is the case for standard nematogens where transitions are
driven by entropic interactions, we expect that the dominant
ordering mechanisms of banana liquid crystals will be associated
with the shape of the molecule and not with electric dipoles. We,
therefore, focus on the mass-moment tensors as the important order
parameters for this problem.  That is, throughout the paper we
will assume that the liquid crystal ordering is driven by steric
interactions and, therefore, that it is the {\em mass}-moment
tensors, rather than charge moments, that are the primary critical
order parameters.

The lowest order mass moment is just the center of mass given, in terms
of the body-fixed coordinate system, by
\begin{mathletters}
\begin{eqnarray}
{\bf R}^\alpha_{\rm cm}&=&{1\over 2m+m_1}\sum_{\mu=1}^3 m_\mu{\bf
R}_{\alpha_\mu}\,\\
&=&\left({m_1\over2m+m_1} a\cos\beta\right){\bbox\nu}_{\alpha,1}\;.
\label{rcm}
\end{eqnarray}
\end{mathletters}
It is natural to define mass moments relative to the center of
mass coordinate ${\bf R}_{\rm cm}^{\alpha}$. Positions of atom
$\mu$ relative to the center of mass are then ${\bf
r}_{\alpha,\mu} = {\bf R}_{\alpha,\mu} - {\bf R}_{\rm
cm}^{\alpha}$. The second mass-moment tensor relative to the
center of mass can be decomposed into a scalar (proportional to
$\delta^{ij}$) and a symmetric, traceless tensor
\begin{equation}
C_{2,\alpha}^{ij}
 = \sum_{\mu=1}^3 m_{\mu}(r_{\alpha,\mu}^ir_{\alpha,\mu}^j-{1\over 3}
r_{\alpha,\mu}^2\delta^{ij}) .
\label{C2alpha}
\end{equation}
The third mass-moment tensor can be decomposed into a vector part,
\begin{equation}
C_{1,\alpha}^i = \sum_{\mu=1}^3 m_{\mu} r_{\alpha,\mu}^2 r_{\alpha,\mu}^i
\end{equation}
and a third-rank symmetric, traceless tensor
\begin{eqnarray}
C_{3, \alpha}^{ijk} & = & \sum_{\mu=1}^3 m_{\mu}\left[r_{\alpha,\mu}^i
r_{\alpha,\mu}^j r_{\alpha,\mu}^k\right.\\
&-&\left.\case{1}{5}r_{\alpha,\mu}^2
(\delta^{ij}r_{\alpha,\mu}^k + \delta^{jk}r_{\alpha,\mu}^i
+ \delta^{ki}r_{\alpha,\mu}^j)\right] .
\label{C3alpha}
\end{eqnarray}
These mass-moment tensors can be expanded in terms of complete sets of
tensors of the appropriate rank formed from the vectors ${\bbox
  \nu}_{\alpha,1}$, ${\bbox\nu}_{\alpha,2}$ and
${\bbox\nu}_{\alpha,3}$:
\begin{mathletters}
\begin{eqnarray}
C_{1,\alpha}^i & = & c_1 \nu_{\alpha,1}^i \\
C_{2,\alpha}^{ij} & = & c_{23} Q_{\alpha,3}^{ij} + c_{22} (Q_{\alpha,1}^{ij}
- Q_{\alpha,2}^{ij} )\\
C_{3,\alpha}^{ijk} & = & c_{31} T_{\alpha,1}^{ijk} + c_{32} T_{\alpha,2}^{ijk},
\end{eqnarray}
\end{mathletters}
where
\begin{equation}
Q_{\alpha,a}^{ij} = \nu_{\alpha,a}^i \nu_{\alpha,a}^j - {
1 \over 3}\delta^{ij} \qquad a=1,2,3 ,
\label{Qalpha}
\end{equation}
\begin{mathletters}
\begin{eqnarray}
T_{\alpha,1}^{ijk} &= &\nu_{\alpha,1}^i
\nu_{\alpha,1}^j\nu_{\alpha,1}^k - {1\over 5}
(\delta^{ij}\nu_{\alpha,1}^k + \delta^{jk}\nu_{\alpha,1}^i +
\delta^{ki}\nu_{\alpha,1}^j) \\
T_{\alpha,2}^{ijk}&= &\nu_{\alpha,3}^i\nu_{\alpha,3}^j\nu_{\alpha,1}^k
+ {\bf\nu}_{\alpha,3}^i\nu_{\alpha,1}^j\nu_{\alpha,3}^k
+ {\bf\nu}_{\alpha,1}^i\nu_{\alpha,3}^j\nu_{\alpha,3}^k\nonumber\\
&-&{1\over 5}(\delta^{ij}\nu_{\alpha,1}^k + \delta^{jk}\nu_{\alpha,1}^i +
\delta^{ki}\nu_{\alpha,1}^j) ,
\end{eqnarray}
\label{Talpha}
\end{mathletters}
and
\begin{mathletters}
\begin{eqnarray}
c_1&=&{2m m_1 a^3\cos\beta(-m_1+2m\cos2\beta)\over (2m+m_1)^2}\;,\\
c_{23}&=&2m a^2\sin^2\beta-{m m_1\over 2m+m_1}a^2\cos^2\beta\;,\\
c_{22}&=&{m m_1\over 2m+m_1}a^2\cos^2\beta\;,\\
c_{31}&=&{2m m_1 (m_1^2+4m^2)\over (2m+m_1)^3}a^3\cos^3\beta\;,\\
c_{32}&=&{4m^2 m_1\over 2m+m_1}a^3\sin^2\beta\cos\beta\;.
\end{eqnarray}
\end{mathletters}
There are only two independent symmetric-traceless molecular parameters
in the set $Q_{\alpha,a}^{ij}$
because the completeness relation,
\begin{equation}
\sum_{a=1}^3 \nu_a^i\nu_a^j =\delta^{ij}\;,
\end{equation}
implies the constraint
\begin{equation}
\sum_{a=1}^3 Q_{\alpha,a}^{ij} = 0\;.
\label{Qconstraint}
\end{equation}
For arbitrary orientation of the axes ${\bbox\nu}_{\alpha,a}$, a given
molecule will in general exhibit $5$ independent symmetric-traceless
tensor mass-moments.  However, because we have chosen these axes to be
symmetry axes of the molecule, there are only two independent tensors.
Similarly, a given molecule will in general be characterized by $7$
independent third-rank mass-moment tensors.  By utilizing $3$
rotational degrees of freedom (e.g., Euler's angles) it is always
possible to choose the axes ${\bbox\nu}_{\alpha,a}$ so that there are
only four independent such tensors.  Our model of a bent-core molecule
is sufficiently simple, that, with our convenient choice of basis
vectors ${\bbox\nu}_{\alpha,a}$, each molecule is characterized by
only $2$ non-vanishing third-rank tensors.

The potential energy of interacting bent-core molecules can be
expressed in terms of the generalized tensors $\nu_{\alpha,1}^i$,
$Q_{\alpha, a}^{ij}$ and $T_{\alpha, a}^{ijk}$ and higher-rank
tensors. In the phenomenological treatment we will pursue, it is
convenient to introduce coarse-grained field versions of these
tensors:
\begin{mathletters}
\begin{eqnarray}
p^i ({\bf x}) & = & {1 \over \rho} \sum_{\alpha}\nu_{\alpha,1}^i
\delta ({\bf x} -{\bf x}_{\alpha} )\;,\label{pi}\\ Q_a^{ij}({\bf
x}) & = &{1 \over \rho}\sum_{\alpha} Q^{ij}_{\alpha,a}\delta ({\bf
x} -{\bf x}_{\alpha} )\;,\label{Qaij}\\ T_a^{ijk}({\bf x}) & = &
{1\over \rho}\sum_{\alpha} T_{\alpha,a}^{ijk}\delta ({\bf x} -{\bf
x}_{\alpha} )\;,
\label{Tijk}
\end{eqnarray}
\label{OPfields}
\end{mathletters}
where $\xv_{\alpha}$ is the position in the lab frame of the
center of mass of molecule $\alpha$ and $\rho$ is the molecular
number density. Thus, a theory for our model bent-core molecules
that includes all tensor order parameters up those of third-rank
would include a single vector order parameter derived from the
third-rank mass-moment tensor, two second-rank tensors and two
third-rank tensors. To simplify our discussion, we will consider
phenomenological theories with only one second-rank tensor, which
we denote $Q^{ij}$, and one third-rank tensor, which we denote
$T^{ijk}$. This, however, is not a restriction on our model,
because if the original theory had all four second- and third-rank
tensors, we could for example integrate out $Q_2^{ij}$ and
$T_1^{ijk}$ order parameters to obtain our model as an effective
theory, depending only on $p^i$, $Q^{ij} \equiv Q_3^{ij}$, and
$T^{ijk}\equiv T_2^{ijk}$ order parameters.

Each of these tensors can be expressed in terms of its components
relative to a space-fixed orthonormal basis $({\bf n}_1, {\bf n}_2,
{\bf n}_3) \equiv ({\bf m}, {\bf l},{\bf n})$, with ${\bf m} \times
{\bf l} = {\bf n}$.  To this end, we introduce second- and third-rank
symmetric-traceless orthonormal basis tensors, $J_{\mu}^{ij}$ and
$I_{\mu}^{ijk}$, that transform, respectively, under $L=2$ and $L=3$
representations of the rotation group in three dimensions:
\begin{mathletters}\begin{eqnarray}
J_1^{ij} & = & \sqrt{3/2}(n^i n^j - {1\over 3} \delta^{ij})\;,\\
J_2^{ij} & = & \sqrt{1/2}(m^i m^j - l^i l^j)\;,\\
J_3^{ij} & = & \sqrt{1/2}(n^i m^j + m^i n^j)\;,\\
J_4^{ij} & = & \sqrt{1/2}(n^i l^j + l^i n^j)\;,\\
J_5^{ij} & = & \sqrt{1/2}(m^i l^j + l^i m^j)\;.
\end{eqnarray}
\label{Js}
\end{mathletters}
and
\begin{mathletters}
\begin{eqnarray}
I_1^{ijk}& = &\sqrt{5/2}[n^i n^j n^k - {1 \over 5} (\delta^{ij} n^k +
\delta^{jk} n^i +\delta^{ki} n^j)]\\
I_2^{ijk} &=& {1 \over 2} (m^i m^j m^k - m^i l^j l^k - m^j l^k l^i -
m^k l^i l^j ) \\
I_3^{ijk} & = & {1 \over 2} (l^i l^j l^k - l^i m^j m^k - l^j m^k m^i
- l^k m^i m^j ) \\
I_4^{ijk}& = & \sqrt{5/12}[m^i n^jn^k + m^j n^k n^i + m^k n^i
n^j\nonumber\\
&-&{1\over 5}(m^i\delta^{jk}+m^j \delta^{ik} + m^k \delta^{ij})]\\
I_5^{ijk} & = & \sqrt{5/12}[l^i n^j n^k + l^j n^k n^i + l^k n^i
n^j\nonumber\\
&-&{1
\over 5}(l^i\delta^{jk}+l^j \delta^{ik} + l^k \delta^{ij})]\\
I_6^{ijk} & = & {1\over \sqrt{6}}[n^i (m^j m^k - l^j l^k) + n^j (m^im^k -
l^i l^k)\nonumber\\
&+& n^k (m^i m^j - l^i l^j)]\\
I_7^{ijk} & = & {1\over\sqrt{6}}(n^i m^j l^k + n^i l^j m^k + m^i l^j n^k +
m^i n^j l^k \nonumber\\
&+& l^i n^j m^k + l^i m^j n^k) .
\end{eqnarray}
\label{Is}
\end{mathletters}
These tensors are normalized so that
\begin{mathletters}
\begin{eqnarray}
\sum_{ij} J_{\mu}^{ij} J_{\mu'}^{ij} & = & \delta_{\mu,\mu'} \\
\sum_{ijk}I_{\mu}^{ijk} I_{\mu'}^{ijk} & = & \delta_{\mu, \mu'} .
\end{eqnarray}
\end{mathletters}
We can now express our order parameters fields, Eq.\ref{OPfields}, in
terms of these bases
\begin{mathletters}
\begin{eqnarray}
p^i & = & \sum_{\mu} p_{\mu} n_{\mu}^i ,\label{Prep} \\
Q^{ij} & = & \sum_{\mu} Q_{\mu} J_{\mu}^{ij},\label{Qrep} \\
T^{ijk} & = & \sum_{\mu} T_{\mu} I_{\mu}^{ijk}.\label{Trep}
\end{eqnarray}
\label{reps}
\end{mathletters}

The space-fixed orthonormal basis $({\bf m}, {\bf l}, {\bf n})$ can be
rotated to diagonalize the tensors $p^i$, $Q^{ij}$, and $T^{ijk}$. In
general, there is no reason why the rotated bases of these three order
parameters should coincide.  We should, therefore, in general
introduce three bases ${({\bf m}_A, {\bf l}_A, {\bf n}_A})$ where
$A\in\{p,Q,T\}$.  Any of these bases are fully specified by three
angles, and we can choose them to eliminate up to three components of
the tensors $Q^{ij}$ and $T^{ijk}$. In particular, we can choose the
two independent angles in ${\bf n}_Q$ to eliminate $Q_3$ and $Q_4$.
This leads to
\begin{eqnarray}
Q^{ij} &=& S(n_Q^i n_Q^j - {1 \over 3} \delta^{ij} ) \nonumber \\
& & + B_1 (m_Q^i m_Q^j - l_Q^i l_Q^j ) + B_2 (m_Q^i l_Q^j + l_Q^i m_Q^j) .
\end{eqnarray}
The independent angle defining the direction of the pair $({\bf m}_Q, {\bf
l}_Q)$ can be used to eliminate either $B_1$ or $B_2$. A similar line of
arguments allows us to choose the basis $({\bf m}_T, {\bf l}_T, {\bf n}_T)$
so that $T_3=T_4=T_5=0$. We can parameterize the four remaining components
of $T^{ijk}$ in terms of an amplitude $T$ and $3$ angles, $\theta_1$,
$\theta_2$, and $\theta_3$ and write
\begin{eqnarray}
T^{ijk}&=&T(\cos{\theta_1}I_1^{ijk}+\sin{\theta_1}\cos{\theta_2}I_2^{ijk}
\nonumber\\
&+&\sin{\theta_1}\sin{\theta_2}\cos{\theta_3}I_6^{ijk}
+\sin{\theta_1}\sin{\theta_2}\sin{\theta_3}I_7^{ijk})\;,\nonumber\\
\label{T4dim}
\end{eqnarray}
in the basis $({\bf m}_T, {\bf l}_T, {\bf n}_T)$. There are other
representations of the general tensor $T^{ijk}$ involving other
sets of four of the tensors $I_{\mu}^{ijk}$ and a different set of
three angles. The representation of Eq.\ (\ref{T4dim}) is the most
useful for out purposes. Finally, we can choose the $p$-basis so
that
\begin{equation}
{\bf p} = p {\bf n}_p .
\end{equation}
In what follows, we will, unless otherwise specified, express all
quantities in the basis $({\bf m}_Q,{\bf l}_Q, {\bf n}_Q)\equiv
({\bf m}, {\bf l}, {\bf n})$ that diagonalizes $Q^{ij}$ with
$B_2=0$. We will then have to worry about the possibility of $7$
independent components of $T^{ijk}$ and three independent
components of $p^i$ in this basis rather than the angles of the
$T$ and $p$ bases relative to those of the $Q$-basis. When
$Q^{ij}$ is zero, we can of course choose the $p$ or the $T$
basis.

\section{Phases and their Symmetries}
\label{PhasesSymmetries}

We have just seen that phases of banana molecules can be
characterized by vector and second- and third-rank-tensor order
parameters. Before developing a Landau field theory for these
order parameters and analyzing it in mean-field theory, we
summarize in this section the various phases and their symmetries
that can arise from these order parameters, and we review possible
phase sequences produced by the simplest version of the Landau
theory.

Before cataloging the possible phases of our model and their
symmetries, we observe that many of these phases can be successfully
described in terms of effective theories that are functions of only
one of the order parameters $p^i$, $Q^{ij}$, or $T^{ijk}$. Some of the
phases, however, particularly those of lower symmetry, require two or
more of these order parameters for their full description.
Furthermore, these order parameters are coupled by rotationally
invariant contributions to the free energy like $p^i p^j Q^{ij}$,
$Q^{ij} T^{ikl}T^{jkl}$, or $p^i Q^{jk} T^{ijk}$, and the existence of
one kind of order induces another.  Thus, for example, a model based
on $T^{ijk}$ alone would miss the fact that the lower-rank tensor
$Q^{ij}$ is automatically induced by $T_2$ order.
\begin{figure*}[bth]
\hspace{-0.32cm}
\begin{tabular}{|lll|}\hline
Phase\hspace{1.cm} & Symmetry\hspace{1.cm} & Order
parameters\hspace{1.cm} \\ \hline\hline
$V$ & C$_{\infty v}$& $p_3$, $S$, $T_1$\hfill\\ \hline
$N$ & D$_{\infty h}$ & $S$ \\ \hline
$N+2$ & D$_{2h}$ & $S$, ${\vec B}_{1,2}$ \\ \hline
$N+3$ & D$_{3h}$ & $S$, ${\vec T}_{2,3}$ \\ \hline
$T$ & T$_d$ & ${\vec T}_{6,7}$ \\ \hline
$N_T$ & D$_{2d}$ & $S$, ${\vec T}_{6,7}$ \\ \hline
$(N_T + 2)^*$ & D$_2$ & $S$, $B_1$, $T_6$, $T_7$\\ \hline
$V+2$ & C$_{2v}$ & $p_3$, $S$, $B_1$, $T_1$, $T_6$; \\
& & or $p_1$, $S$, $B_1$, $T_2$, $T_4$\\ \hline
$V+3$ & C$_{3v}$ & $p_3$, $S$, $T_1$, ${\vec T}_{2,3}$\\ \hline
$(V_T + 2)^*$ & C$_2$ & $p_3$, $S$, $B_1$, $T_1$, $T_6$, $T_7$;\\
& & or $p_1$, $S$, $B_1$, $T_2$, $T_4$, $T_5$\\ \hline 
$N+V$ & C$_{1h}$ & $p_1$, $p_3$, $S$, $B_1$, $Q_3$,\\ & &
$T_1$, $T_2$, $T_4$, $T_6$ \\ \hline
\end{tabular}

{Table 1: Anisotropic liquid phases of banana-shaped molecules,
their symmetries in the Schoenflies notation, and their
nonvanishing order parameters. The notation ${\vec B}_{1,2}$, etc.
is explained in the text. Some of the phases, such as the $N+V$
phase can be characterized by other sets of symmetry equivalent
order parameters, involving, for example, linear combinations of
$p_1$ and $p_2$ rather than $p_1$ alone.}
\end{figure*}

Table 1 lists the phases we consider, their symmetries, and the
nonvanishing order parameters that characterize them. This list
includes phases with all symmetries that can be constructed from
the order parameters $p^i$, $Q^{ij}$, and $T^{ijk}$ except for the
lowest-symmetry phase with $C_1$ symmetry\cite{commentN+V*,SmC_G}, 
which we do not
consider. All other point-group symmetries including cubic,
icosahedral, simple tetrahedral ($T$), and even lower symmetries
such as $S_2$, $C_3$, and $C_{2h}$ cannot be characterized without
the introduction of 4th- or higher-rank tensor order parameters.
As is customary, we denote the Isotropic phase by $I$ and the
Nematic phase by $N$. The $N$ phase has D$_{\infty h}$ symmetry,
and it is completely characterized within the space of $p^i$,
$Q^{ij}$, and $T^{ijk}$ by the single uniaxial order parameter
$S$. In general, the $N$ phase will also have nonvanishing
components of all even rank tensors (explicitly induced through
$\mbox{Tr}(Q^n T_{2n})$ coupling), but we will ignore them,
focusing on the nontrivial order parameters of rank $3$ or less
that actually drive the ordering transitions. There is a phase
with vector or, equivalently, C$_{\infty v}$ symmetry, which we
denote by $V$. The predominant order parameter of this phase is
the vector ${\bf p}$, which we take to be along ${\bf n}$ (i.e.,
nonzero $p_3$). Once $p_3$ orders, it explicitly induces $S$ and
$T_1$ order parameters, through the $p^i p^j p^k T^{ijk}$ and $p^i
p^j Q^{ij}$ couplings, respectively.

There are a number of phases in which anisotropy develops in the
plane perpendicular to ${\bf n}$ or ${\bf p}$. As we will find in
Sec.\ \ref{PhaseTransitions}, phases that break uniaxial symmetry
(isotropy of the transverse plane), will exhibit $O(2)$ invariance
corresponding to internal rotation within pairs of order
parameters. For convenience we will collectively refer to these
pairs as: $\vec{p}_{1,2}\equiv(p_1,p_2)$,
$\vec{B}_{1,2}\equiv(B_1,B_2)$, $\vec{T}_{2,3}\equiv(T_2,T_3)$,
$\vec{T}_{4,5}\equiv(T_4,T_5)$, $\vec{T}_{6,7}\equiv(T_6,T_7)$.
Also, following Toner\cite{TonerN+r}, we denote the phases with
$r$-fold anisotropy (or equivalently $r$-atic order) in the plane
perpendicular to ${\bf n}$ by $N+r$ and those with similar
anisotropy perpendicular to the vector axis ${\pv}$ by $V+r$.
There is the standard biaxial nematic or $N+2$ phase with D$_{2h}$
symmetry and $S$ and $\vec{B}_{1,2}$ order. There is an $N+3$
phase with D$_{3h}$ symmetry and nonvanishing $S$ and
$\vec{T}_{2,3}$ order. The $V+2$ (equivalent to the $N+1$ phase)
and $V+3$ phases have C$_{2v}$ and C$_{3v}$ symmetry,
respectively. The $V+3$ phase develops from the $N+3$ phase by
developing vector order along the ${\bf n}$ axis. It, therefore,
has $p_3$ and $T_1$ order in addition to the $S$ and
$\vec{T}_{2,3}$ order of the $N+3$ phase.  The $V+2$ phase has one
$2$-fold axis and two perpendicular reflection planes. Which order
parameters describe this phase depends on whether the vector order
${\bf p}$ lies along ${\bf n}$ (which diagonalizes $Q^{ij}$) or
perpendicular to ${\bf n}$, within the ${\bf m}$-${\bf l}$ plane.
If ${\bf p}$ is parallel to ${\bf n}$, the $V+2$ phase is
characterized by nonvanishing $p_3$, $S$, $B_1$, $T_1$, and $T_6$
or by a symmetry equivalent set such as $p_3$, $S$, $B_2$, $T_1$,
and $T_7$, where it is understood here that $T_7$ is zero if $B_1$
is nonzero and $T_6$ is zero if $B_2$ is nonzero (otherwise
spontaneous chirality develops, as discussed below); if ${\bf p}$
is perpendicular to ${\bf n}$, then it is characterized by
nonvanishing $\vec{p}_{1,2}$, $S$, $B_1$, $T_2$, and $T_4$ or
symmetry-equivalent order parameters.

\begin{figure}[bth]
\centering
\setlength{\unitlength}{1mm}
\begin{picture}(150,165)(0,0)
\put(-47,-85){\begin{picture}(150,0)(0,0)
\includegraphics{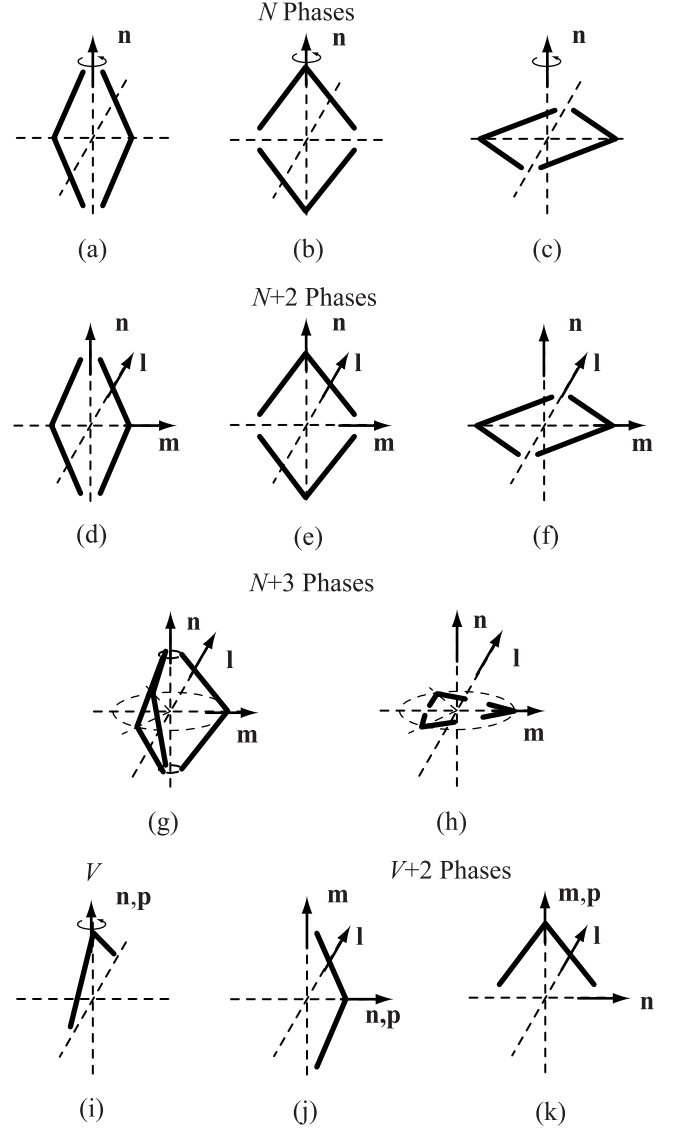}
\end{picture}}
\end{picture}
\caption{Schematic representation of the $N$, $N+2$, $N+3$, $V$,
  and $V+2$ phases.  Three versions, (a), (b), and (c), of the $N$
  phases are depicted with respective predominant alignment of
  ${\bbox{\nu}_3}$, ${\bbox{\nu_1}}$, and ${\bbox{\nu_2}}$ along $\bf
  n$, specifying the direction of the principal axis of $Q^{ij}$ with
  the largest eigenvalue. The $N+2$ phases, (d), (e), and (f), are
  obtained, respectively, from the $N$ phases (a), (b), and (c) by
  restricting rotations in the plane perpendicular to ${\bf n}$ to
  have two-fold symmetry, whereas the $N+3$ phases, (g) and (h), are
  obtained by restricting these rotations to have a three-fold
  symmetry. In the uniaxial $V$ phase (i), the molecular ${\bbox \nu}_1$ aligns
  on average along ${\bf p} || {\bf n}$, sampling equally all
  orientations about the ${\bf p}$ axis.  The
  $V+2$ phase can be produced either by introducing biaxial order
  perpendicular to ${\bf p}$ and ${\bf n}$ into the $V$ phase (j) or
  by introducing vector order into the $N+2$ phase by aligning ${\bf
    p}$ along ${\bf m}$ (k). }
\label{phases_fig}
\end{figure}

Schematic representations of the $N$, $N+2$, $N+3$, $V$, and $V+2$
phases derived from bent-core molecules are shown in Fig.\
\ref{phases_fig}. The distribution of molecular angles in the
$V+3$ phase is difficult to depict in the format of Fig.\
\ref{phases_fig}. Because of the symmetry of the bent-core
molecule under ${\bbox \nu}_3 \rightarrow - {\bbox \nu}_3$, it is
impossible to produce $V+3$ symmetry if the molecular ${\bbox
  \nu}_1$ axis is rigidly aligned along ${\bf p}$.  To produce such
3-fold symmetry, it is necessary for the molecular ${\bbox \nu}_1$
axis to be tilted away from the the ${\bf p}$ axis and for its
projection onto the plane perpendicular to ${\bf p}$ to have
three-fold symmetry.

A comment about how $\vec{T}_{2,3}$ order describes $N+3$ (and $V+3$)
symmetry is useful.  $r$-atic order is generally described by an order
parameter of the form $\langle e^{i r \phi} \rangle$, where $\phi$ is
the angle between a molecular axis in the ${\bf m}$-${\bf l}$ plane
and the ${\bf m}$ axis.  To represent $\vec{T}_{2,3}$ order in this
way, we introduce the circular basis vectors
\begin{equation}
{\bf e}_{\pm} = {1 \over \sqrt{2}} ( {\bf m} \pm i {\bf l} )
\label{circular}
\end{equation}
satisfying
\begin{equation}
{\bf e}_+ \cdot {\bf e}_- = 1 , \qquad {\bf e}_+ \cdot {\bf e}_+ = 0,
\qquad {\bf e}_- \cdot {\bf e}_-= 0 ,
\end{equation}
and reexpress $I_2^{ijk}$ and $I_3^{ijk}$ as
\begin{equation}
I_2^{ijk} = {1\over\sqrt{2}}(I_+^{ijk} + I_-^{ijk}), \qquad
I_3^{ijk} = {-i\over\sqrt{2}}(I_+^{ijk} - I_-^{ijk}) .
\end{equation}

Then $T_2 I_2^{ijk} + T_3 I_3^{ijk} = T_+ I_+^{ijk} + T_- I_-^{ijk}$, where
\begin{equation}
T_{\pm} = {1\over\sqrt{2}}(T_2 \mp i T_3) = {\bf e}_{\pm}^i {\bf e}_{\pm}^j
{\bf e}_{\pm}^k T^{ijk} . \label{T+-}
\end{equation}
If ${\bbox \nu}_3$ is aligned along ${\bf n}$, then $T_{\pm}
= \langle e^{\pm i 3 \phi} \rangle$ where $\phi$ is the angle between
${\bbox \nu}_1$ and $\bf m$.  When ${\bbox \nu}_3$ is not 
aligned along ${\bf n}$, the situation is similar though more
complicated.  Thus, nonzero $\vec{T}_{2,3}$ describes {\em triadic}
order in the plane perpendicular to ${\bf n}$.
We will also briefly encounter even lower symmetry phases, in
which, in contrast to the $N+r$ and $V+r$ phases discussed above,
the additional order develops in a plane that is {\em not}
perpendicular to the established nematic or vector axis. One
prominent example is a phase in which the nematic and polar orders
are neither parallel nor perpendicular. We will refer to this
$C_{1h}$-symmetry phase as $N+V$, emphasizing its distinction from
the $N+1$ ($\equiv V+2$) phase, discussed above.  Although $N+V$
phase (and its $N+r+V$ analogs) is quite unlikely to develop in
the liquid state, such order can quite naturally appear in the
smectic-$C$ environment, where the additional axis is defined by
the smectic layer-normal $\bf{N}$.
\begin{figure}[bth]
\centering \setlength{\unitlength}{1mm}
\begin{picture}(60,55)(0,0)
\put(-33,-75){\begin{picture}(150,0)(0,0)
\includegraphics{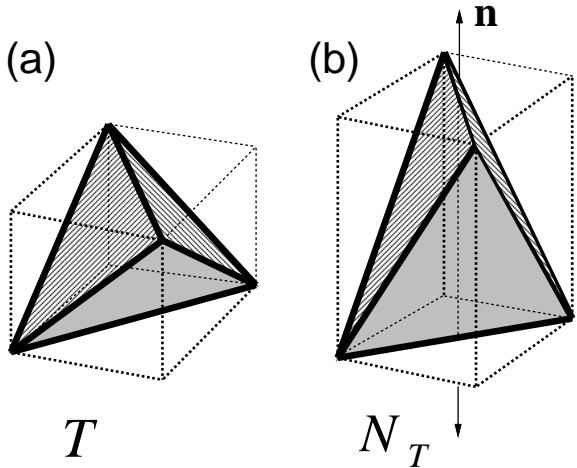}
\end{picture}}
\end{picture}
\caption{(a) A tetrahedron, that exhibits symmetry identical to that of
  the Tetrahedratic $T$ phase, with all three axes of the cube, ${\bf n},
  {\bf m}, {\bf l}$ equivalent. The $T$ phase can be visualized
  as being composed of banana molecule tetrahedral complexes on
  average decorating edges of randomly positioned but
  orientationally-ordered tetrahedra as shown in Fig.\
  \ref{tetrafig}(a).  (b) A tetrahedron uniaxially
  distorted along the ${\bf n}$ axis, exhibiting symmetry of the $N_T$
  phase, that is distinguished from the $T$ phase by the nonzero
  nematic order parameter $S$. A depiction of this phase in terms of
  banana molecules is shown in Fig.\ \ref{tetrafig}(b)}
\label{T-N_Tphases}
\end{figure}

\begin{figure}[bth]
\centerline{\epsfbox{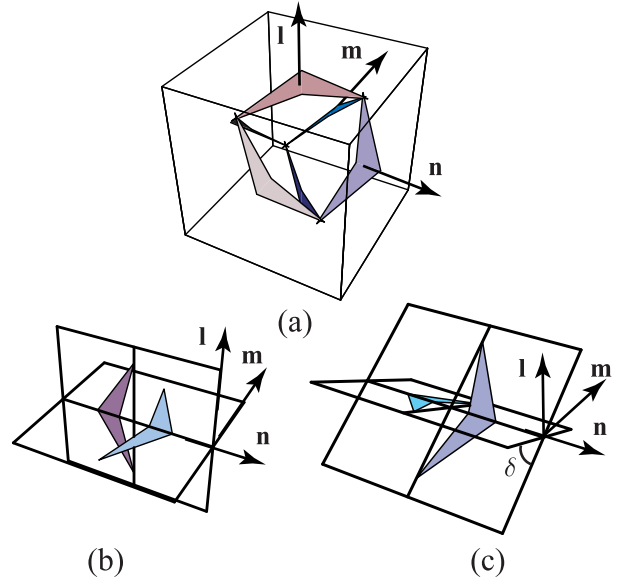}}
\caption{Schematic representation in terms of banana molecules
  of (a) the $T$ phase, (b) the $N_T$ phase, and (c) the $(N_T + 2)^*$ phase.
  In the $T$-phase,
  bent-core molecules align locally with their ${\bbox \nu_3}$ axes
  aligned on average along the six edges of a tetrahedron and their
  ${\bbox \nu_1}$ axes aligned parallel to the normals $\pm {\bf m}$,
  $\pm {\bf l}$, and $\pm {\bf n}$ to these edges.  Opposite edges are
  (say with normals along $\bf n$ and $- {\bf n}$) are orthogonal so
  that molecules aligned along opposite edges have perpendicular
  ${\bbox \nu}_3$ axes. The $N_T$ phase is obtained from the $T$
  phase by a uniaxial distortion along one of the cubic axes as
  shown in Fig.\ \ref{T-N_Tphases} to favor
  one pair of crossed bent-core molecules over the other two
  orthogonal pairs. Note the invariance of the $N_T$ phase under
  the 4-fold improper rotation $S_4: {\bf m} \rightarrow {\bf l}, {\bf
  l} \rightarrow - {\bf m}, {\bf n} \rightarrow - {\bf n}$. The
  chiral nonpolar $(N_T + 2)^*$ phase is obtained from the $N_T$ phase by
  rotating the two molecular planes away from $90^o$ to an angle
  $0<\delta<\pi/2$ to remove the $S_4$ symmetry element, as also 
  illustrated in Fig.\ref{N_Tphases}.}
\label{tetrafig}
\end{figure}

There is only one phase in which $T^{ijk}$ has a nonvanishing
component and in which {\em both} $p^i$ and $Q^{ij}$ are zero.
This phase, which we denote by $T$ and call tetrahedratic, has
tetrahedral symmetry and is invariant under all 24 operations of
the tetrahedral group T$_d$\cite{Tinkham}. It is characterized by
an arbitrary nonvanishing linear combination of $T_6$ and $T_7$,
i.e., by the $\vec{T}_{6,7}$ order parameter, and is illustrated
in Figs.\ \ref{T-N_Tphases}(a) and \ref{tetrafig}(a).

\begin{figure}[bth]
\centering
\setlength{\unitlength}{1mm}
\begin{picture}(60,155)(0,0)
\put(-33,0){\begin{picture}(150,0)(0,0)
\includegraphics{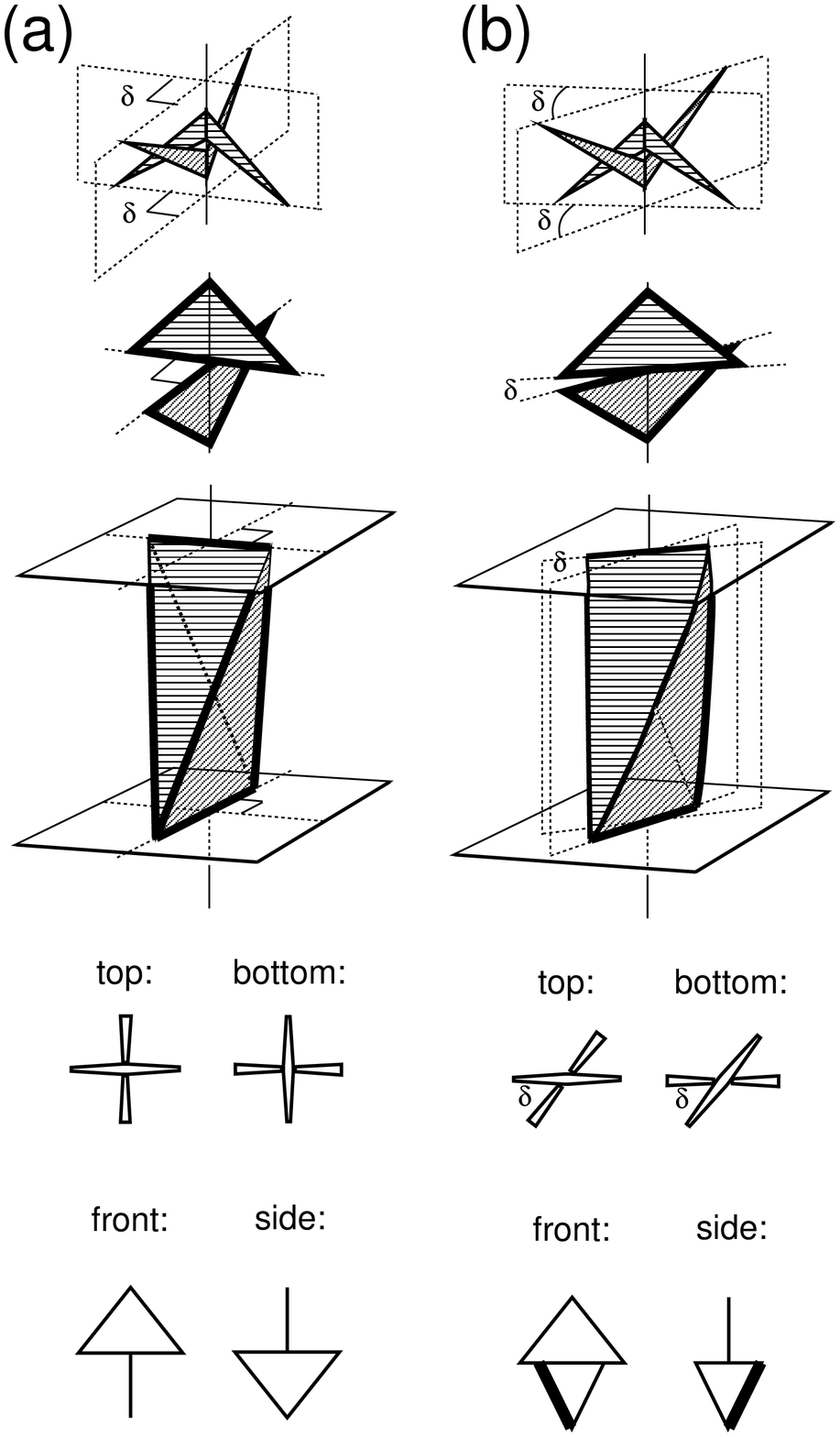}
\end{picture}}
\end{picture}
\caption{Various representations of the achiral  $N_T$ (a) and the chiral
  $(N_T+2)^*$ (b) phases. The two phases are distinguished by their
  opening angle $\delta$.  In the achiral $N_T$ phase, $\delta =
  \pi/2$, whereas in the chiral $(N_T + 2)^*$ phase,
  $0<\delta<\pi/2$.}
\label{N_Tphases}
\end{figure}

A uniaxial distortion along one of the $3$ two-fold tetrahedral
axes reduces the T$_d$ symmetry of the tetrahedratic phase down to
D$_{2d}$ symmetry. We denote the resulting {\em nonpolar} phase
with this symmetry by $N_T$. It is characterized by nonvanishing
nematic $S$ and $\vec{T}_{6,7}$ order parameters, but it is
neither a uniaxial nor a biaxial nematic, with $\vec{T}_{6,7}$
breaking the isotropy of the plane transverse to ${\bf n}$, as
illustrated in Figs.\ \ref{T-N_Tphases}(b), \ref{tetrafig}(b), and
\ref{N_Tphases}(a).

The $V+2$ phase with $C_{2v}$ symmetry naturally emerges from the
$N_T$ phase through the development of longitudinal polar order
${\bf p}=p_3{\bf n}$ along the existing nematic axis. As we will
see in Sec.\ref{PhaseTransitions}, once $p_3$ develops in the
presence of $\vec{T}_{6,7}$, a biaxial order $\vec{B}_{1,2}$ with
principal axes {\em parallel} to those of $\vec{T}_{6,7}^{ijk}n^i$
is explicitly induced. Our final two phases, which we denote by
$(N_T + 2)^*$ and $(V_T + 2)^*$, respectively, have D$_2$ and
C$_2$ symmetry. They are unique in that they are spontaneously
{\em chiral} phases.  The {\em nonpolar chiral} $(N_T + 2)^*$
phase depicted in Figs.\ \ref{tetrafig}(c) and \ref{N_Tphases}(b)
is formed from the nonpolar achiral $N_T$ phase by the development
of biaxial $\vec{B}_{1,2}$ order (but in contrast to the polar
achiral $V+2$ phase) with principle axes {\em rotated exactly} by
$\pi/4$ relative to those of the $\vec{T}_{6,7}^{ijk}n^i$ order
parameter, which characterizes the $N_T$ phase. More concretely,
for the choice of the basis ${\bf m}$-${\bf l}$, such that $N_T$
is {\em exclusively} described by nonvanishing $S$ and $T_7$ order
parameters, the polar achiral $V+2$ and the nonpolar chiral $(N_T
+ 2)^*$ phases emerge when $B_2$ and $B_1$, respectively, order;
equivalently, if it is the nonzero $S$ and $T_6$ that are {\em
exclusively} used to describe the $N_T$ phase, then the roles of
$B_2$ and $B_1$ are reversed and transitions to $V+2$ and $(N_T +
2)^*$ take place when $B_1$ and $B_2$, respectively, become
nonzero. The polar chiral $(V_T + 2)^*$ phase emerges from the
nonpolar chiral $(N_T +2)^*$ phase via development of polar order
$p_3$ along the existing nematic ${\bf n}$ axis. 

\begin{figure}[bth]
\centerline{\epsfbox{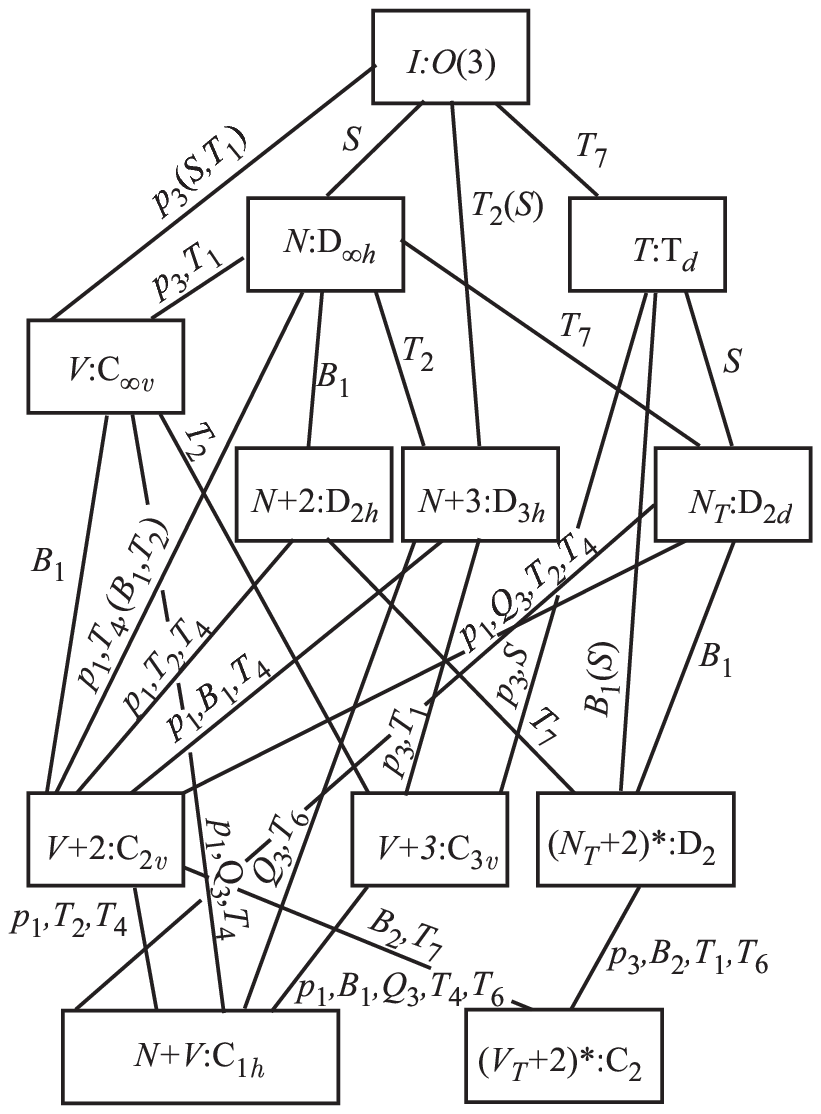}}
\vspace{30pt}
\caption{A flowchart of phase transitions between liquid crystal
  phases illustrated in Fig.\ \ref{phases_fig}. Order parameters, which
  become nonzero at each of the transitions and their symmetry groups
  are indicated. For transitions that we have studied in detail, we have
  also indicated the secondary (explicitly induced by nonlinear couplings) 
  order parameters by placing them in parenthesis.}
\label{flowchart}
\end{figure}

Alternatively, a
transition to it can also take place from the polar achiral $V+2$
phase by spontaneously breaking chiral symmetry via development
of biaxial $\vec{B}_{1,2}$ order with principle axes {\em rotated
exactly} by $\pi/4$ from those of $\vec{T}_{6,7}^{ijk}n^i$ order
parameter. Not surprisingly, in all the polar phases the $T_1$
order parameter is also explicitly induced. Since both the $(N_T
+2)^*$ and $V_T +2)^*$ phases are chiral, their ground-state
configurations will exhibit spatial modulations like those of
cholesteric and blue phases of chiral mesogens.

Given the {\em achirality} of the bent-core molecules, the
transitions from $N_T$ to $(N_T + 2)^*$ and $V+2$ to $(V_T + 2)^*$
are ones in which chiral symmetry is broken {\em spontaneously},
and they are, therefore, relevant to the physics of chiral banana
phases.\cite{Link}
Figure \ref{flowchart} summarizes the phases we treat and the
symmetry-lowering transitions among them that {\em can} take
place, i.e., allowed by symmetry (as opposed to energetic)
considerations. The fact that C$_{2v}$ is a subgroup of C$_{\infty
v}$ (${\rm C}_{2v} \subset {\rm C}_{\infty v}$) implies that there
can be a $V \rightarrow V+2$ symmetry-lowering transition.  The
subgroup structure ${\rm D}_2 \subset {\rm D}_{2d} \subset {\rm
T}_d\subset O(3)$ and ${\rm C}_{2v} \subset {\rm D}_{3h} \subset
O(3)$, where $O(3)$ is the full orthogonal group including
inversions in three dimensions, imply, respectively that the $I
\rightarrow T \rightarrow N_T \rightarrow (N_T + 2)^*$ and $I
\rightarrow N+3 \rightarrow V+2$ phase sequences are possible.
Other phase sequences shown in Fig.\ \ref{flowchart} follow from
similar group-theoretic arguments, and are supported by the
detailed analysis of the Landau mean-field theory given in
Sec.\ref{PhaseTransitions}.

\section{Construction of a Landau Field Theory}
\label{LandauTheory}

Having identified the possible spatially homogeneous but anisotropic
phases\cite{comment_anisotropy} of systems described by first, second,
and third-rank tensor order parameters, we now turn to the study of
phase transitions among them. To this end we begin by constructing a
Landau free energy that will describe transitions from the Isotropic
phase. The appropriate Landau free energy functional is a rotationally
invariant power-series expansion in the order parameters $p^i$,
$Q^{ij}$, and $T^{ijk}$. The most general Landau free-energy density
is produced by sums of scalars formed from the tensors $p^i$,
$Q^{ij}$, and $T^{ijk}$.  It can be decomposed as
\begin{equation}
f = f_p + f_Q + f_T + f_{pQ} + f_{pT} + f_{QT} + f_{pQT} + f_{Q^2 T^2},
\end{equation}
where $f_p$, $f_Q$, $f_T$ are, respectively, the Landau energies
for independent vector, second-rank, and third-rank tensor order
parameters and the other energies are couplings between these
order parameters. The vector energy $f_p$ is given by the standard
$O(N=3)$ model:
\begin{equation}
f_p = \case{1}{2} K_p (\partial_j p^i)(\partial_j p^i)
+ \case{1}{2} r_p p^i p^i + u_p (p^i p^i)^2\;.
\label{fp}
\end{equation}
The purely $2$nd-rank-tensor part of $f$ is the well-known Landau
energy for Isotropic-Nematic
transition\cite{deGennesProst,ChaikinLubensky}, given by
\begin{eqnarray}
f_Q &=& \case{1}{2}K_Q(\partial_k Q^{ij}\partial_k Q^{ij})+
{1\over2} r_Q Q^{ij} Q^{ij}\nonumber\\
& & - w_Q Q^{ij} Q^{jk} Q^{ki}+u_Q (Q^{ij}Q^{ij})^2\;,
\label{fQ}
\end{eqnarray}
and the purely $3$rd-rank-tensor part of $f$ is given by
\begin{eqnarray}
f_T &=& \case{1}{2}K_T(\partial_l T^{ijk}\partial_l T^{ijk})+
\case{1}{2} r_T T^{ijk} T^{ijk} \label{fT}\\
& & + u_T(T^{ijk}
T^{ijk})^2 + v_T T^{i_1 i_2 i_3} T^{i_1 i_4 i_5} T^{i_2 i_4 i_6}
T^{i_3 i_5 i_6}\;. \nonumber
\end{eqnarray}
In the above expressions we have suppressed the position
dependence of order parameters, have used an Einstein convention
for the repeated indices, and have left out the dipolar-like
(``space-spin'' coupling) gradient terms, $\partial_i
Q^{ik}\partial_j Q^{jk}$, and $\partial_{i_1} T^{i_1 j k}
\partial_{i_2}T^{i_2 j k}$, that couple internal indices of
$Q^{ij}$ and $T^{ijk}$ to that of the spatial coordinate ${\bf
x}$. Although this last simplification might modify the asymptotic
nature of the phase transitions, it obviously will {\em not}
affect our mean-field discussions, valid outside of a (typically)
narrow critical region. The parameters $r_p \sim T - T_p$, $r_Q
\sim T- T_Q$, and $r_T \sim T - T_T$ vanish at the temperatures
$T_p$, $T_Q$ and $T_T$, which are determined predominantly by the
interaction potential between molecules, that characterize
mean-field limits of metastability. In writing down quartic
nonlinearities in $f_{Q}$ and $f_{T}$, we have used the
non-obvious
relations (valid in 3 dimensions) %
\begin{mathletters}
\begin{eqnarray}
{1\over2}(Q^{ij}Q^{ji})^2&=&Q^{ij}Q^{jk}Q^{kl}Q^{li}\;,\\
{1\over2}(T^{ijk} T^{ijk})^2&=&T^{i j_1 k_1}T^{l j_1 k_1} T^{i j_2
k_2} T^{l j_2 k_2}\nonumber\\
&& + T^{i_1 i_2 i_3} T^{i_1 i_4 i_5} T^{i_2 i_4 i_6}T^{i_3 i_5 i_6}\;,
\label{quartics}
\end{eqnarray}
\end{mathletters}
to reduce the number of quartic couplings in $f_{Q}$ from two to one
and in $f_{T}$ from three to two.  The lowest-order contributions to
the coupling energies are
\begin{mathletters}
\begin{eqnarray}
f_{pQ} & = & - w_{pQ}\ p^i p^j Q^{ij}\label{fpQ} \\
f_{pT}& = & - w_{pT}\ p^i p^j p^k T^{ijk}\label{fpT}\\
f_{QT}& = &- w_{QT}\ Q^{i_1 i_2} T^{i_1 j k} T^{i_2 j k}\label{fQT} \\
f_{pQT}&=& -w_{pQT}\ p^i Q^{jk} T^{ijk}\label{fpQT}\\
f_{Q^2T^2}^{(1)}& = &- w_{1}\ Q^{i_1 l} Q^{i_2 l}
T^{i_1 j k} T^{i_2 j k}\label{f1} \\
f_{Q^2T^2}^{(2)}& = &- w_{2}\ Q^{i_1 j_1} Q^{i_2 j_2}
T^{i_1 j_1 k} T^{i_2 j_2 k}\label{f2} \\
f_{Q^2T^2}^{(3)}& = &- w_{3}\ Q^{i_1 i_2} Q^{j_1 j_2} T^{i_1 j_1
k} T^{i_2 j_2 k}
\label{f3}\;,
\end{eqnarray}
\label{fc}
\end{mathletters}
where we have decomposed $f_{Q^2 T^2}$ as $\sum_{n=1}^3f_{Q^2
T^2}^{(n)}$. The term $p^i Q^{jk}T^{ijk}$ deserves special
attention. If the product $Q^{jk}T^{ijk}$ is nonzero this term
will induce vector $p^i$ order. Thus, it is possible to have a
transition from the $N$ phase that appears to be driven by
$T^{ijk}$ but which nonetheless develops vector order.  In other
words, a model expressed in terms of $Q^{ij}$ and $T^{ijk}$, only,
would miss the development of vector order, which by itself is
unlikely to order in a realistic liquid crystal.

As usual, the average properties are computed by integrating over
order parameter configurations with a Boltzmann weight with an
effective Hamiltonian ${\cal H} = \int d^3 x f$
\begin{equation}
\langle{\cal O}\rangle={1\over{\cal Z}}\int{\cal D}p^i {\cal D}
Q^{ij} {\cal D} T^{ijk}\; {\cal O}(p^i,Q^{ij},T^{ijk})\;e^{-{\cal
H}/k_BT}\;,
\end{equation}
where ${\cal O}(p^i,Q^{ij},T^{ijk})$ is a function of the order
parameters $p^i$, $Q^{ij}$ and $T^{ijk}$ and $\cal Z$ is the
partition function.

\section{Phase Transitions}
\label{PhaseTransitions}

In the preceding two sections, we defined order parameters, we
identified possible rotationally anisotropic, but spatially
homogeneous thermodynamic phases of bent-core molecules, and we
constructed a Landau theory to describe phase transitions among
these phases. In this section, we will show how each of the phase
transition sequences depicted in Fig.\ \ref{flowchart} arises in
mean-field theory. We will organize our discussion by considering
sequentially each symmetry-lowering transition from each of the
phases in Fig.\ \ref{flowchart}. Thus, we will first discuss
transitions from the $I$ phase to the $V$, $N$, $N+3$, and $T$
phases. We will then study transitions from each of these phases,
that is from the $V$, $N$, $N+3$, and $T$ phases into
lower-symmetry phases and so on until the lowest-symmetry $V+3$,
$(V_T+2)^*$, and $N+V$ phases are reached.

In our discussion of transitions out of various partially ordered
phases, it will be useful to display explicitly the part of the
free energy functional $f$ that couples the order parameters
$p^i$, $Q^{ij}$, and $T^{ijk}$, which identify our phases. This
coupling energy part, which we collectively call $f_c$, arises
from the $Q^3$ and $Q^4$ parts of $f_Q$, from the $T^4$ parts of
$f_T$, and from $f_{pQ}$, $f_{pT}$, $f_{QT}$, $f_{pQT}$,
$f_{Q^2T^2}^{(1)}$, $f_{Q^2T^2}^{(2)}$, and $f_{Q^2T^2}^{(3)}$. It
determines those order parameters that are coupled at harmonic
order when long-range order has been established in a subset of
order parameters, and it is, therefore, essential in establishing
the nature of the phases and phase diagrams for our system.

\subsection{Transitions from the Isotropic phase}

The Isotropic phase is the phase with the highest ($O(3)$)
symmetry. A symmetry-lowering transition to the $V$ phase with
vector symmetry takes place with the development of ${\bf p}$
order and one to the $N$ phase with the development of uniaxial
$Q^{ij}$ order. As Figs.\ \ref{flowchart} and \ref{PhaseDiagram}
indicate, the development of $T^{ijk}$ order in the Isotropic
phase $I$, can lead to two distinct phases: the tetrahedral phase
$T$ with ${\vec T}_{6,7}$ order and the $N+3$ phase with uniaxial
$Q^{ij}$ order in addition to ${\vec T}_{2,3}$ order. At long length scales,
small fluctuations within the Isotropic phase are described by a
harmonic free energy density
\begin{equation}
{\tilde f}^{(I)}= \case{1}{2} ( r_p p^i p^i + r_Q Q^{ij} Q^{ij} + r_T T^{ijk}
T^{ijk} ) .
\end{equation}
Thus, which of the fields $p^i$,$Q^{ij}$, or $T^{ijk}$ first becomes
unstable is determined by which of the set of parameters ${\cal S}_I =
\{r_p, r_Q, r_T\}$ first passes through zero.  Some of the transitions
from the $I$ phase are, however, first-order, and which of the
possible transitions actually takes place depends on higher-order
terms in the free energy.  We will thus consider each transition
separately.

\subsubsection{$I\rightarrow V$ transition}

\begin{figure}[bth]
\centering
\setlength{\unitlength}{1mm}
\begin{picture}(150,45)(0,0)
\put(-20,-57){\begin{picture}(150,0)(0,0)
\includegraphics{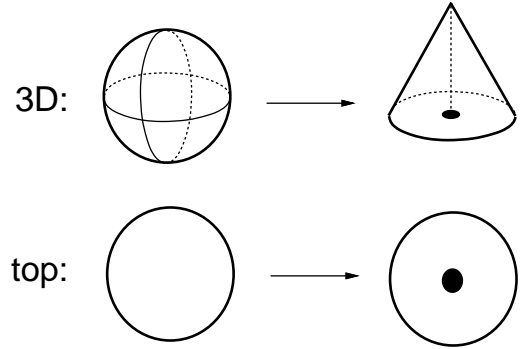}
\end{picture}}
\end{picture}
\caption{A schematic for the $I\rightarrow V$ transition.}
\label{I-Vtransition}
\end{figure}

The $I-V$ transition is driven by the development of ${\bf p}$ order.
Since ${\bf p}$ is a vector, this transition is in the well-known
$O(3)$ universality class; it can be described in terms of an
effective theory involving ${\bf p}$ only.  Below this second-order
transition, we can take ${\bf p}$ to point along the $3$ direction
with
\begin{equation}
p_3 \sim |r_p - r_{pc}|^{\beta_{O(3)}}\;,
\end{equation}
with $\beta_{O(3)} \approx 0.366$, where $r_{pc}$ is the value of
$r_p$ at the critical point.  Once ${\bf p}$ develops, it drives both
$S$ and $T_1$ order via the interactions
\begin{eqnarray}
f_{pQ} & = & - \case{2}{3} w_{pQ} p_3^2 S \\
f_{pT} & = & - \sqrt{\case{2}{5}} w_{pQT} p_3^3 T_1 \;,
\end{eqnarray}
so that in mean-field theory\cite{mft_exponent}
\begin{mathletters}
\begin{eqnarray}
S &\sim& |r_p - r_{pc}|^{2 \beta_{O(3)}},\\
T_1 &\sim& |r_p - r_{pc}|^{3 \beta_{O(3)}},
\end{eqnarray}
\end{mathletters}
for $r_p < r_{pc}$ in the $V$ phase.

\subsubsection{$I\rightarrow N$ transition}
\begin{figure}[bth]
\centering
\setlength{\unitlength}{1mm}
\begin{picture}(150,55)(0,0)
\put(-20,-57){\begin{picture}(150,0)(0,0)
\includegraphics{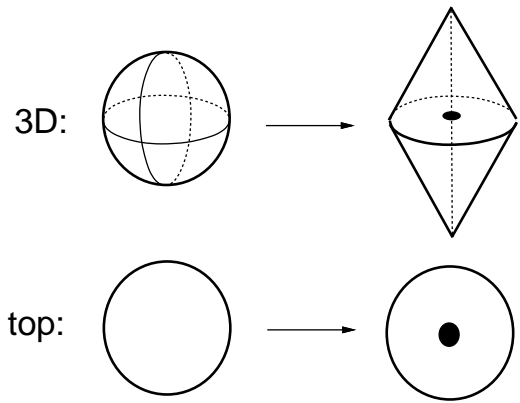}
\end{picture}}
\end{picture}
\caption{A schematic for the $I\rightarrow N$ transition.}
\label{I-Ntransition}
\end{figure}

The $I-N$ transition is driven by the development of $Q^{ij}$
order from the Isotropic phase. There are no couplings that
explicitly drive either $p^i$ or $T^{ijk}$ order once $Q^{ij}$
order develops. Consequently, this well-studied
transition\cite{deGennesProst,ChaikinLubensky}, which is described
completely by the $f_Q$ part of the free energy density, is
generically first-order and in mean-field theory takes place at
$r_Q = w_Q^2/12 u_Q$. A direct transition from the isotropic to
the biaxial nematic ($N+2$) phase is also possible. Since it is
fairly complicated and has been treated in detail\cite{alban}, we
will not consider it further here.

\subsubsection{Transitions from $I$ driven by $T^{ijk}$}
\label{TbeforeQ}
Transitions from the $I$ phase involving the development of
third-rank tensor\cite{fel1,fel2} order are more complex than the
other transitions from the $I$ phase we have considered. The many
degrees of freedom in the $T^{ijk}$ tensor lead to the possibility
of two distinct transitions, the $I\rightarrow T$ and the $I
\rightarrow N+3$ transition\cite{fel2}. Since it is fluctuations
in $T^{ijk}$ that drive these transitions, the noncritical degrees
of freedom, $p^i$ and $Q^{ij}$ can safely be integrated out to
produce an effective theory involving only $T^{ijk}$ whose free
energy is identical in form to Eq.\ (\ref{fT}). This is the energy
that we will use to study transitions from the $I$ phase involving
$T^{ijk}$ order. We will, however, have cause to return to the
more general theory in our discussion of the $I\rightarrow N+3$
transition. There are two important things to note about the free
energy $f_T$ in Eq.\ (\ref{fT}). First, in contrast to $f_Q$, this
energy has no odd-order invariants because none can be formed with
a third-rank tensor, $T^{ijk}$. Second, there are two fourth order
invariants\cite{fel2}, which as we shall see, compete in the
determination of the symmetry of the order parameter that develops
from the Isotropic phase. In the limit of vanishing $v_T$
coupling, $f_T$ is invariant under the operations of the group
$O(7)$, as can be seen by reexpressing $f_T$ with $v_T = 0$ as
\begin{equation}
f_{T}^{O(7)}={1\over2}r_T|{\vec T}|^2 + u_T|{\vec
T}|^4\;,\label{ftO7}\\
\end{equation}
where ${\vec T}$ is a $7$-dimensional vector with components
$T_\mu$ defined by Eq.\ (\ref{Trep}). Because the underlying
$O(3)$ symmetry of our system is lower than $O(7)$, with $T^{ijk}$
forming its $7$-dimensional irreducible representation, it is not
surprising that the full free energy is {\em not} $O(7)$
invariant. The $v_T$ quartic term, explicitly breaks the $O(7)$
symmetry and determines which of the $7$ $T_\mu$ irreducible
components of $T^{ijk}$ order at the transition from the Isotropic
phase.

To determine which components order, it is convenient to use the
alternative representation of $T^{ijk}$ given by Eq.\
(\ref{T4dim}). In this representation we have
\begin{equation}
f_{T}={1\over2}r_T T^2 + u_T T^4 +
f_{v_T}\;,\label{Ht07Scalar}
\end{equation}
with
\begin{eqnarray}
f_{v_T}&=&{v_T T^4\over150}\Bigl[9\cos^4{\theta_1}
+{15\over2}\sin^2{2\theta_1}(1+2\cos{2\theta_2})\nonumber\\
&+&25\sin^4{\theta_1}\sin^4{\theta_2}\Bigr]\;,\label{Hvt}
\end{eqnarray}
breaking the $O(7)$ invariance of $f_T$. $f_{v_T}$ (and thus
$f_T$) has an $O(2)$ invariance, which, through our choice of
parameterization of $T^{ijk}$, Eq.\ (\ref{T4dim}), is manifested
by $f_{v_T}$'s being independent of $\theta_3$.  Thus, finding the
minimum-energy state requires the minimization of
$f_T(\theta_1,\theta_2)$ over two rather than three angles.

\begin{center}
{\it (i.) $I\rightarrow N+3$ transition}
\end{center}

\begin{figure}[bth]
\centering
\setlength{\unitlength}{1mm}
\begin{picture}(150,53)(0,0)
\put(-20,-57){\begin{picture}(150,0)(0,0)
\includegraphics{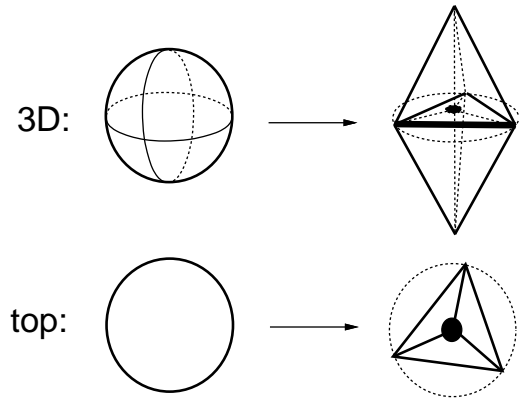}
\end{picture}}
\end{picture}
\caption{A schematic for the $I\rightarrow N+3$ transition.}
\label{I-N3transition}
\end{figure}

Following standard analysis, for $v_T>0$ we find that the global
minimum of $f_{v_T}(\theta_1,\theta_2)$ at fixed $T$ is given by
\begin{mathletters}
\begin{eqnarray}
\theta_1^{\rm min}&=&{\pi\over2}\;,\label{theta1a}\\
\theta_2^{\rm min}&=&0\;,\label{theta2a}
\end{eqnarray}
\label{theta12a}
\end{mathletters}
which corresponds to a state with
\begin{equation}
T^{ijk}_{\rm min}=T_2 I_2^{ijk}\;, \label{vT_positive_n}
\end{equation}
i.e., a state with a planar triadic order [see Eqs.\
(\ref{circular}) to (\ref{T+-})], here chosen to lie in the ${\bf
m}-{\bf l}$ plane. A rotation within this plane shows that for a
more general choice [than that defined by the representation
choice, Eq.(\ref{T4dim})] of ${\bf m}$ and ${\bf l}$ axes relative
to the molecular body axes, such triadic planar order is
described by an arbitrary linear combination of the $I_2^{ijk}$
and $I_3^{ijk}$ tensors, corresponding to nonvanishing $T_2$ and
$T_3$ order parameters, i.e., a nonvanishing $\vec{T}_{2,3}$.

Since this triadic order defines a plane that brings with it a normal
invariant under reflection, it necessarily induces uniaxial nematic
order, $Q^{ij} = S(n^i n^j -
\case{1}{3}
\delta^{ij} )$ with ${\bf n}$ along the normal. To lowest order,
the development of $S$ is brought about by the $f_{QT}$ coupling
of Eq.\ (\ref{fQT}),
\begin{equation}
f_{QT}={1\over3}w_{QT} T_2^2 S\; ,
\end{equation}
which leads to the expected uniaxial nematic order with
\begin{equation}
S=-{T_2^2\over 2r_Q}\;.
\end{equation}
Hence the state for $v_T>0$ is the $N+3$ phase in which the
nematic and triadic order, transverse to the nematic axis,
coexist. From the point of view of symmetry it is equivalent to a
liquid of orientationally ordered equilateral triangles with
aligned normals.

Another solution that minimizes the energy
$f_T(\theta_1,\theta_2)$ and that is degenerate with the state
described by the solution in Eq.\ (\ref{theta12a}) is
\begin{mathletters}
\begin{eqnarray}
\theta_1^{\rm min}&=&\arccos{\sqrt{5/8}}\;,\label{theta1b}\\
\theta_2^{\rm min}&=&{\pi\over2}\;.\label{theta2b}
\end{eqnarray}
\label{theta12b}
\end{mathletters}
It corresponds to a state
\begin{equation}
T^{ijk}={T\over\sqrt{8}}[\sqrt{5}I_1^{ijk}+\sqrt{3}I_6^{ijk}]\;,
\label{vT_positive_m}
\end{equation}
which is equivalent to $T^{ijk}_{\rm min}=T I_2^{ijk}$, Eq.\
(\ref{vT_positive_n}) after ${\bf n}$ and ${\bf m}$ are interchanged.
Clearly then this solution also represents the $N+3$ phase, but with
the nematic axis along ${\bf m}$ rather than ${\bf n}$, and the triadic
order in the ${\bf n}-{\bf l}$ plane.

\begin{center}
{\it (ii.) $I\rightarrow T$ transition}
\end{center}

\begin{figure}[bth]
\centering
\setlength{\unitlength}{1mm}
\begin{picture}(150,30)(0,0)
\put(-20,-57){\begin{picture}(150,0)(0,0)
\includegraphics{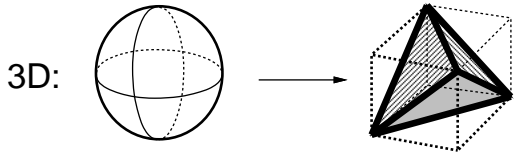}
\end{picture}}
\end{picture}
\caption{A schematic for the $I\rightarrow T$ transition.}
\label{I-Ttransition}
\end{figure}

When $v_T<0$ the global minimum of $f_{v_T}(\theta_1,\theta_2)$ at
fixed $T$ is given by
\begin{mathletters}
\begin{eqnarray}
\theta_1^{\rm min}&=&{\pi\over2}\;,\label{theta1c}\\
\theta_2^{\rm min}&=&{\pi\over2}\;,\label{theta2c}
\end{eqnarray}
\label{theta12c}
\end{mathletters}
which corresponds to  the state with
\begin{equation}
T^{ijk}_{\rm min}=T_6 I_6^{ijk}\;,\label{vT_negative_n}
\end{equation}
which is invariant under the operations of the tetrahedral group
$T_d$. As can be seen in Fig.\ \ref{T-N_Tphases}, the group $T_d$
has three $C_2$ axes coinciding with the axes of the cube, four
$C_3$ axes coinciding with the body diagonals of the cube, six
reflection planes passing through each edge and bisecting the
opposite edge of the tetrahedron, and four $S_4$ improper rotation
axes corresponding to the axes bisecting (four) sets of two
opposite edges of the tetrahedron. Because this state lacks ${\bf
n}\rightarrow-{\bf n}$ symmetry, no nematic or any other order is
induced by the coupling free energy $f_c$, Eq.\ (\ref{fc}).
Because of its tetrahedral symmetry and because only the $T_6$
component of $T^{ijk}$ is nonzero, we identify this state with the
$T$ phase illustrated schematically in Fig.\ \ref{tetrafig}(a).
Since $f_{v_T}$ is independent of $\theta_3$, an arbitrary linear
combination of $T_6$ and $T_7$, rather than $T_6$ alone, will in
general become nonzero at the $I \rightarrow T$ transition.

\begin{figure}[bth]
\centering
\setlength{\unitlength}{1mm}
\begin{picture}(150,80)(0,0)
\put(-20,-57){\begin{picture}(150,0)(0,0)
\includegraphics{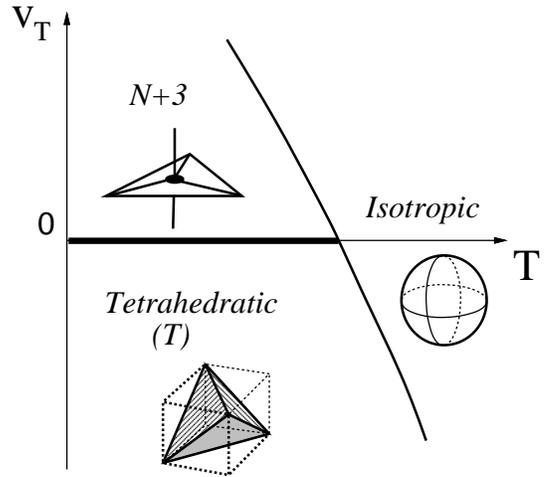}
\end{picture}}
\end{picture}
\caption{A portion of a phase diagram for banana-shaped liquid crystal,
  illustrating two possible transitions out of the Isotropic phase.
  For $v_T>0$, the transition is $I\rightarrow N+3$, and for $v_T<0$
  it is the $I\rightarrow T$ transition. Although in mean-field theory
  these transitions are continuous, we expect thermal fluctuations to
  drive them first-order. Lowering temperature along a finely-tuned
  $v_T=0$ curve, we expect a continuous transition in the $O(7)$
  universality class.}
\label{PhaseDiagram}
\end{figure}

For $v_T<0$ we also find another solution that minimizes the energy $f_T$ and
is degenerate with the state in Eq.\ (\ref{theta12c}). It is given by 
\begin{mathletters} 
\begin{eqnarray} 
\theta_1^{\rm min}&=&\arccos{\sqrt{5/9}}\;,\label{theta1d}\\ 
\theta_2^{\rm min}&=&0\;,\label{theta2d} 
\end{eqnarray} 
\label{theta12d} 
\end{mathletters} 
and corresponds to a state with 
\begin{equation}
T^{ijk}={T_6\over3}[\sqrt{5}I_1^{ijk} +2I_2^{ijk}]\;. \label{vT_positive_m2}
\end{equation} 
However, it can easily be shown that this solution is
equivalent to $T^{ijk}_{\rm min}=T I_6^{ijk}$, after a rotation around the
$\bf l$ axis: 
\begin{mathletters} 
\begin{eqnarray} 
{\bf n}&\rightarrow&\sqrt{1/3}{\bf n}+\sqrt{2/3}{\bf m}\;,\label{rotation1}\\ 
{\bf m}&\rightarrow&-\sqrt{2/3}{\bf n}+\sqrt{1/3}{\bf m}\;.\label{rotation2}
\end{eqnarray} 
\label{rotation12} 
\end{mathletters} 
It, thus, also corresponds to the $T$-phase with pure $T_6$ order in a rotated
coordinate system. The corresponding phase diagram that graphically summarizes
phase transitions outlined above is given in Fig.\ \ref{PhaseDiagram}.  

For $v_T=0$, it is clear from Eqs.\ (\ref{Ht07Scalar}) and
(\ref{Hvt}) that the transition from the $I$ phase is in the
$O(7)$ universality class, and it is to a state that {\em
spontaneously} breaks $O(7)$ symmetry by picking out a particular
direction in the $O(7)$ symmetric space (a point on a
$7$-dimensional sphere) for the vector $T_\mu$ to point in.
Clearly, as we have seen above, the $v_T$ coupling is relevant in
the ordered phase and drives the resulting state toward $N+3$
phase for $v_T>0$ and toward $T$ for $v_T<0$.

We have also investigated the stability of the $O(7)$ symmetric
{\em transition} (with $v_T=0$) to a finite value of the $v_T$
symmetry breaking interaction\cite{ITpaper}. In a renormalization
group calculation, just below the upper-critical dimension $d=4$,
we find that in the presence of thermal fluctuations the $v_T$
coupling always drives this transition first-order. This is
analogous to the similar phenomena known in magnetic systems in
which cubic crystal fields drive the $O(3)$ transition of
hypothetical isotropic magnets first-order\cite{Rudnick,Brezin}.

\subsection{Transitions from the $N$ phase}

As illustrated in the flowchart of Fig.\ \ref{flowchart}, there
are $5$ symmetry-reducing transitions from the nematic phase.
These are the $N\rightarrow V$, $N\rightarrow N+2$, $N\rightarrow
N+3$, $N\rightarrow V+2$, and $N\rightarrow N_T$ transitions, all
of which we will discuss in detail below. To determine which
transitions will occur for a given set of phenomenological
parameters, we focus on the part of the full free energy density,
${\tilde f}^{(N)}$, describing harmonic fluctuations about the
nematic phase with nonvanishing $S$. This free energy, ${\tilde
f}^{(N)}$, is determined by the harmonic parts of the energies
$f_p$, $f_Q$, and $f_T$, and by the coupling terms $f_c$.  The
most important contributions to $f_c$ come from $f_{pQ}$, and
$f_{pQT}$, which can be easily evaluated.  The harmonic free
energy ${\tilde f}^{(N)}$ can be expressed as a sum of five
independent parts:
\begin{equation}
{\tilde f}^{(N)} = {\tilde f}_{p_3,T_1}^{(N)}+{\tilde
f}_{B_{1,2}}^{(N)}+{\tilde f}_{T_{2,3}}^{(N)}+{\tilde
f}_{p_{1,2},T_{4,5}}^{(N)}+{\tilde f}_{T_{6,7}}^{(N)},
\end{equation}
where
\begin{mathletters}
\begin{eqnarray}
{\tilde f}_{p_3,T_1}^{(N)} & = &
\case{1}{2} \tilde{r}_{p_3}^{(N)} p_3^2 +
\case{1}{2} \tilde{r}_{T_1}^{(N)} T_1^2 +
\tilde{\alpha}_{p_1,T_3}^{(N)} p_3 T_1 \\
{\tilde f}_{B_{1,2}}^{(N)} & = &
\case{1}{2}\tilde{r}_{B_{1,2}}^{(N)} (B_1^2 + B_2^2) \\
{\tilde f}_{T_{2,3}}^{(N)} & = &
\case{1}{2}\tilde{r}_{T_{2,3}}^{(N)} (T_2^2 + T_3^2) \\
{\tilde f}_{p_{1,2},T_{4,5}}^{(N)} & = &
\case{1}{2}\tilde{r}_{p_{1,2}}^{(N)} (p_1^2 + p_2^2) +
\case{1}{2} \tilde{r}_{T_{4,5}}^{(N)}(T_4^2 + T_5^2)\nonumber\\
&+&\tilde{\alpha}_{p_{1,2},T_{4,5}}^{(N)}(p_1 T_4 + p_2 T_5)\label{fp1T4} \\
 {\tilde f}_{T_{6,7}}^{(N)} & = &
\case{1}{2}\tilde{r}_{T_{6,7}}^{(N)} (T_6^2 + T_7^2)
\end{eqnarray}
\end{mathletters}
with
\begin{eqnarray}
\hspace*{-2cm}\tilde{r}_{p_3}^{(N)}& = & r_p - \case{4}{3} w_{pQ}
S,\nonumber\\
\tilde{r}_{T_1}^{(N)} & = & r_T - \case{8}{15} w_{QT} S -
(\case{28}{45}w_1+\case{4}{5}w_2+\case{2}{9}w_3)S^2,\nonumber\\
\tilde{\alpha}_{p_3,T_1}^{(N)}& = &
-\sqrt{\case{2}{5}}w_{pQT}S,\nonumber\\
\tilde{r}_{B_{1,2}}^{(N)} & = & 2r_Q  + 4w_Q S + \case{16}{3}u_Q
S^2,\nonumber\\
\tilde{r}_{T_{2,3}}^{(N)} & = & r_T +\case{2}{3} w_{QT}S -
\case{2}{9}(w_1+w_3)S^2,\nonumber\\
\tilde{r}_{p_{1,2}}^{(N)} & = & r_p + \case{2}{3}w_{pQ} S,\nonumber\\
\tilde{r}_{T_{4,5}}^{(N)} & = & r_T - \case{2}{5} w_{QT} S -
(\case{26}{45}w_1+\case{8}{15}w_2+\case{2}{45}w_3)S^2,\nonumber\\
\tilde{\alpha}_{p_{1,2},T_{4,5}}^{(N)}& = &
-\case{2}{\sqrt{15}}w_{pQT}S,\nonumber\\
\tilde{r}_{T_{6,7}}^{(N)} & = & r_T - \case{2}{9}(2w_1 -
w_3)S^2,
\label{rs2}
\end{eqnarray}
Within mean-field theory, the $N$ phase becomes unstable to the
development of biaxial order (characterized by a linear
combination of $B_1$ and $B_2$), of triaxial order (characterized
by a linear combination of $T_2$ and $T_3$), and of $N_T$ order
(characterized by a linear combination of $T_6$ and $T_7$), when
$\tilde{r}_{B_{1,2}}^{(N)}$\cite{B12positive},
$\tilde{r}_{T_{2,3}}^{(N)}$, and $\tilde{r}_{T_{6,7}}^{(N)}$,
respectively, pass through zero.  The $N$ phase becomes unstable
to longitudinal vector order (characterized by coupled $p_3-T_1$
order parameters) and to transverse vector order (characterized by
coupled $p_1-T_4$ order parameters), when the determinants,
\begin{mathletters}
\begin{eqnarray}
\Delta_{p_3,T_1}^{(N)} &=&
\tilde{r}_{p_3}^{(N)} \tilde{r}_{T_1}^{(N)} -
\left(\tilde{\alpha}_{p_3,T_1}^{(N)}\right)^2 \\
\Delta_{p_{1,2},T_{4,5}}^{(N)}  &= &
\tilde{r}_{p_{1,2}}^{(N)} \tilde{r}_{T_{4,5}}^{(N)} -
\left(\tilde{\alpha}_{p_{1,2},T_{4,5}}^{(N)}\right)^2
\end{eqnarray}
\end{mathletters}
respectively, pass through zero. The nature of transitions out of the
$N$ phase will be determined by which member of the set, ${\cal S}_N =
\{\Delta_{p_3,T_1}^{(N)}, \tilde{r}_{B_{1,2}}^{(N)},
\tilde{r}_{T_{2,3}}^{(N)}, \tilde{r}_{T_{6,7}}^{(N)},
\Delta_{p_{1,2},T_{4,5}}^{(N)}\}$, first passes through zero on
lowering the temperature.

\subsubsection{$N\rightarrow V$ transition}

\begin{figure}[bth]
\centering
\setlength{\unitlength}{1mm}
\begin{picture}(150,55)(0,0)
\put(-20,-57){\begin{picture}(150,0)(0,0)
\includegraphics{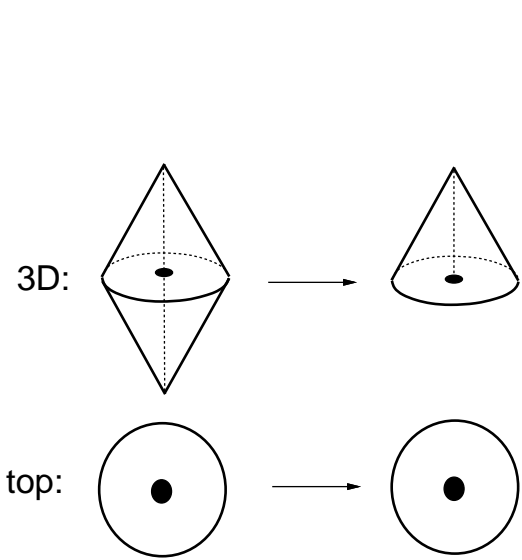}
\end{picture}}
\end{picture}
\caption{A schematic for the $N\rightarrow V$ transition.}
\label{N-Vtransition}
\end{figure}

The $N \rightarrow V$ transition is signaled by the development of
vector order along the unique direction ${\bf n}$ of the $N$ phase,
i.e., by the development of $p_3$ and $T_1$ order.  Thus, this
transition occurs if $\Delta_{p_3,T_1}^{(N)}$ is the first of the set
${\cal S}_N$ to become zero. The relative sign of $p_3$ and $T_1$ is
fixed by the eigenfunction associated with the smallest eigenvalue of
the matrix defined by ${\tilde f}^{(N)}_{p_3,T_1}$.  The overall sign
is, however, arbitrary. Since it is the associated ${\cal Z}_2$
symmetry (with $V$ pointing along or antiparallel to ${\bf n}$) that
is broken, the $N \rightarrow V$ transition is in the well-studied
Ising universality class with coupled order parameters
\begin{mathletters}
\begin{eqnarray}
p_3 &\sim& |\Delta_{p_3,T_1}^{(N)}-\Delta_{p_3,T_1}^{(N)c}|^{\beta_{\rm Ising}},\\
T_1 &\sim&
|\Delta_{p_3,T_1}^{(N)}-\Delta_{p_3,T_1}^{(N)c}|^{\beta_{\rm
Ising}},
\end{eqnarray}
\end{mathletters}
growing for $\Delta_{p_3,T_1}^{(N)}<\Delta_{p_3,T_1}^{(N)c}$. Once
$p_3$ and $T_1$ become nonzero, they do {\em not} force the
development of any other order, and the $V$ phase is completely
characterized by $p_3$, $S$, and $T_1$ order parameters, as
illustrated in Figs.\ \ref{phases_fig} and \ref{flowchart}.

\subsubsection{$N\rightarrow N+3$ transition}

\begin{figure}[bth]
\centering
\setlength{\unitlength}{1mm}
\begin{picture}(150,55)(0,0)
\put(-20,-57){\begin{picture}(150,0)(0,0)
\includegraphics{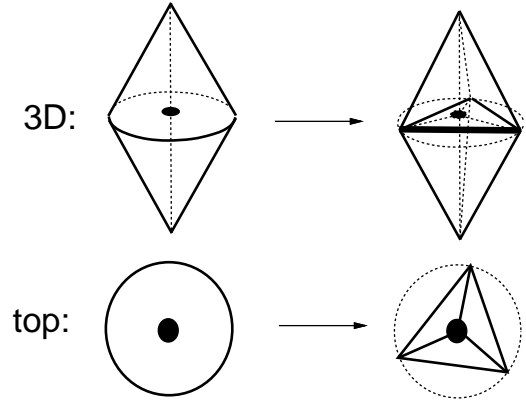}
\end{picture}}
\end{picture}
\caption{A schematic for the $N\rightarrow N+3$ transition.}
\label{N-N3ransition}
\end{figure}

The $N+3$ phase develops out of the $N$ phase with the appearance of a
linear combination of $T_2$ and $T_3$ order.  As discussed in Eqs.\
(\ref{circular}) to (\ref{T+-}), $T_2$ and $T_3$ define a
two-dimensional representation of the group of rotations perpendicular
to ${\bf n}$ and describe triadic order in the plane perpendicular to
${\bf n}$.  Since the $N$ phase is invariant with respect to
arbitrary rotations about ${\bf n}$, the free energy of the $N$ phase
is a function only of the rotationally invariant combinations $T_2^2 +
T_3^2=|{\vec T}_{2,3}|^2$.  Thus, the $N\rightarrow N+3$ 
transition is in the well-known $XY$ universality class.  Within 
mean-field theory, this transition occurs when $\tilde{r}_{T_{2,3}}^{(N)}$ 
is the first in the set ${\cal S}_N$ to pass through zero on cooling.  
${\vec T}_{2,3}$ order drives no other order, and the $N+3$ phase is 
completely characterized by $S$ and ${\vec T}_{2,3}$ with
\begin{eqnarray}
T_2\propto T_3&\sim&
|\tilde{r}_{T_{2,3}}^{(N)}-\tilde{r}_{T_{2,3}c}^{(N)}|^{\beta_{XY}},
\end{eqnarray}
for $\tilde{r}_{T_{2,3}}<\tilde{r}_{T_{2,3}c}^{(N)}$.

\subsubsection{$N\rightarrow N+1$ ($\equiv V+2$) transition}

\begin{figure}[bth]
\centering
\setlength{\unitlength}{1mm}
\begin{picture}(150,55)(0,0)
\put(-20,-57){\begin{picture}(150,0)(0,0)
\includegraphics{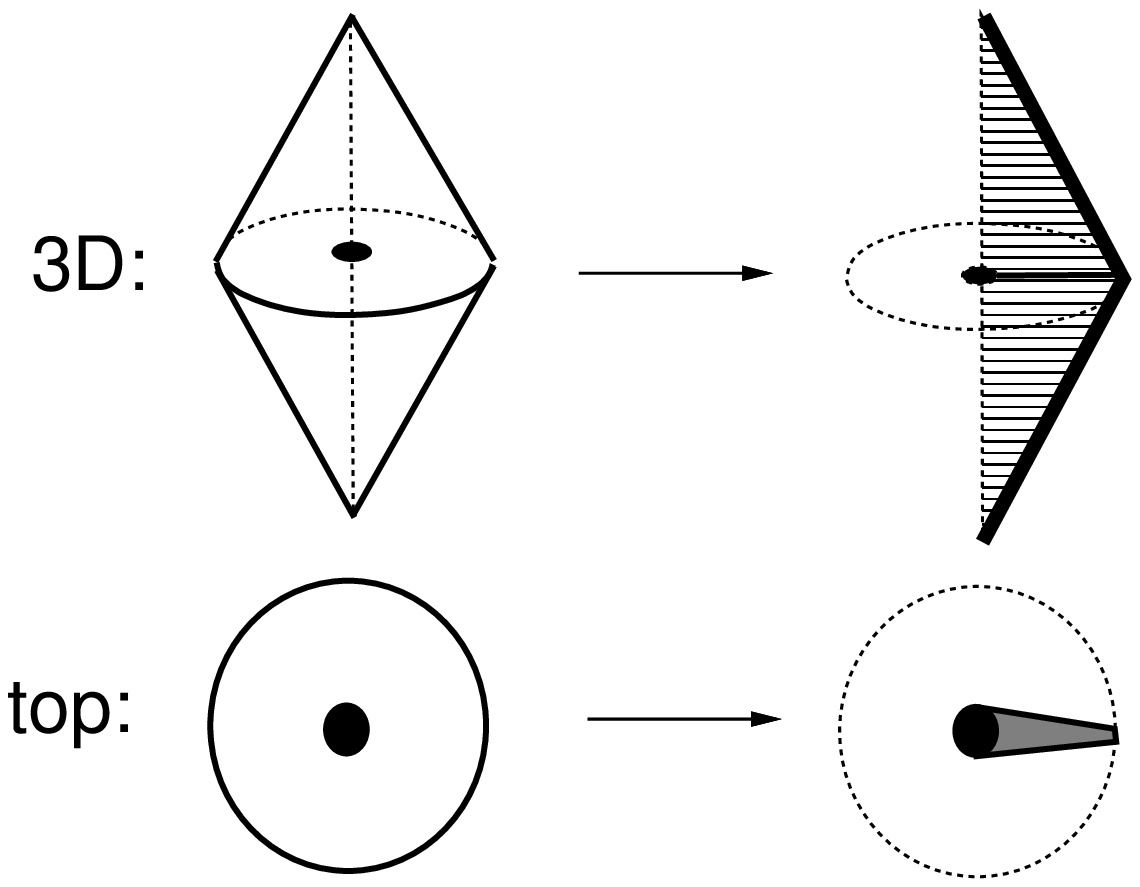}
\end{picture}}
\end{picture}
\caption{A schematic for the $N\rightarrow N+1$ transition.}
\label{N-N1ransition}
\end{figure}

The $V+2$ phase is distinguished from the $N$ phase by the
existence of a vector order described by ${\bf p}$ in the plane
{\em perpendicular} to ${\bf n}$. We can arbitrarily choose ${\bf
p}$ to be along ${\bf m}$ so that $p_1$ is nonzero. Since there is
already uniaxial order in the $N$ phase, $T_4$ order has the same
symmetry in the $N$ phase as does $p_1$ order, and not
surprisingly $p_1$ and $T_4$ (as are generically $p_2$ and $T_5$) are coupled
in ${\tilde f}^{(N)}$, Eq.\ (\ref{fp1T4}). The invariance of the
$N$ phase with respect to arbitrary rotations about ${\bf n}$
implies that the energy of the $N$ phase must be a function of
rotationally invariant combinations $p_1^2 + p_2^2=|{\vec p}_{1,2}|^2$, 
$T_4^2 + T_5^2=|{\vec T}_{4,5}|^2$ and $p_1 T_4 + p_2 T_5={\vec
p}_{1,2}\cdot {\vec T}_{4,5}$ as the harmonic free energy
${\tilde f}_{p_{1,2}, T_{4,5}}^{(N)}$ in Eq.\ (\ref{fp1T4}) is.
The $N\rightarrow V+2$ transition is thus in the $XY$ universality
class. It occurs in mean-field theory when
$\Delta_{p_{1,2},T_{4,5}}^{(N)}$ is the first of the set ${\cal
S}_N$ to pass through zero upon cooling. The order parameters
$p_1$ and $T_4$ (and $p_2$ and $T_5$ related to them by a rotation
in the $m-l$ plane) will drive the $N\rightarrow V+2$ transition,
both growing continuously from zero as
\begin{eqnarray}
p_1\propto T_4&\sim&
|\Delta_{p_1,T_4}^{(N)}-\Delta_{p_1,T_4}^{(N)c}|^{\beta_{XY}},
\end{eqnarray}
for $\Delta_{p_1,T_4}^{(N)}<\Delta_{p_1,T_4}^{(N)c}$.  Once these
order parameters become nonzero, however, they pick out a
direction in the plane perpendicular to ${\bf n}$ that drives the
development of a non-vanishing biaxial order $B_1$ via the
$f_{pQ}$ and $f_{QT}$ coupling free energies, as is clear from
Fig.\ \ref{N-N1ransition}. In mean-field theory, $B_1 \sim p_1^2$
and $T_2 \sim p_1^3$.  Below the critical dimension $d_c=4$,
however, potentials in the coupling energies are relevant,
and\cite{mft_exponent,Aharony}
\begin{equation}
B_1 \sim p_1^{\sigma_2}\qquad T_2 \sim p_1^{\sigma_3} ,
\end{equation}
where $\sigma_n = n + x_n n(n-1)$ with $x_n$ only weakly dependent
on $n$.

To emphasize the secondary role that $B_1$ and $T_2$ order
parameters play at the $N\rightarrow V+2$ transition, in Fig.\
(\ref{flowchart}), $B_1$ and $T_2$ are placed in parenthesis along
the $N\rightarrow V+2$ line. Therefore, the $V+2$ phase is reached
from the $N$ phase (characterized by a finite value of $S$) via
the well-studied, second order $XY$ transition upon the
development of $p_1$, $B_1$, $T_2$ and $T_4$ order parameters or a
set that can be obtained from this one by an $XY$-rotation.

\subsubsection{$N\rightarrow N+2$ transition}

\begin{figure}[bth]
\centering
\setlength{\unitlength}{1mm}
\begin{picture}(150,55)(0,0)
\put(-20,-57){\begin{picture}(150,0)(0,0)
\includegraphics{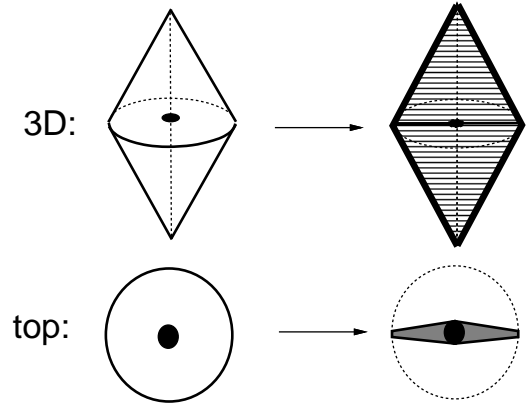}
\end{picture}}
\end{picture}
\caption{A schematic for the $N\rightarrow N+2$ transition.}
\label{N-N2ransition}
\end{figure}

The development of nonvanishing biaxial order parameter
$\vec{B}_{1,2}=(B_1,B_2)$ converts the uniaxial $N$ phase to the
$N+2$ phase. The biaxial order parameter is a rank-$2$
symmetric-traceless tensor, which, because of its confinement to
the two-dimensional plane perpendicular to the uniaxial axis ${\bf
n}$, is equivalent to a complex order parameter forming an
irreducible $L=2$ representation of the $U(1)$ group. The
$N\rightarrow N+2$ transition is thus in the XY universality
class.  In mean-field theory, it takes place when
$\tilde{r}_{B_{1,2}}$ is the first of the set ${\cal S}_N$ to pass
through zero and more generically, in the presence of thermal
fluctuations, we expect,
\begin{eqnarray}
B_1\propto B_2&\sim&
|\tilde{r}_{B_{1,2}}^{(N)}-\tilde{r}_{B_{1,2} c}^{(N)}|^{\beta_{XY}},
\end{eqnarray}
for $\tilde{r}_{B_{1,2}}^{(N)}<\tilde{r}_{B_{1,2} c}^{(N)}$.  Biaxial
order forces no $p^i$ or $T^{ijk}$ order, so the $N+2$ phase is fully
characterized by nematic order parameter $S$ and an arbitrary linear
combination of $B_1$ and $B_2$ biaxial order parameters, i.e.,
by $\vec{B}_{1,2}$.

\subsubsection{$N\rightarrow N_T$ transition}

\begin{figure}[bth]
\centering
\setlength{\unitlength}{1mm}
\begin{picture}(150,55)(0,0)
\put(-20,-57){\begin{picture}(150,0)(0,0)
\includegraphics{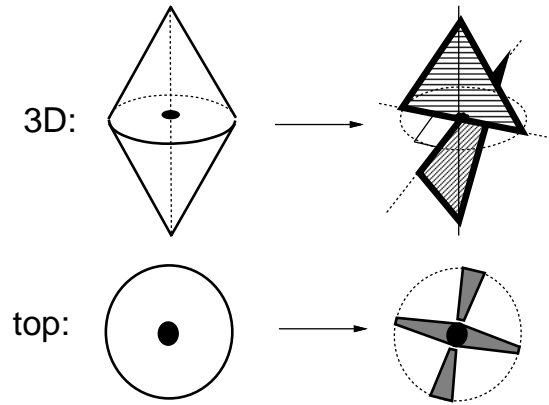}
\end{picture}}
\end{picture}
\caption{A schematic for the $N\rightarrow N_T$ transition.}
\label{N-N_Ttransition}
\end{figure}

Finally, the $N_T$ phase is distinguished from the $N$ phase by
the development of an arbitrary linear combination of the $T_6$
and $T_7$ i.e., of the $\vec{T}_{6,7}$ order.  Since such an order
parameter picks out a single direction within the {\em isotropic}
plane perpendicular to ${\bf n}$, this transition, like the
$N\rightarrow N+2$, $N\rightarrow N+3$, and $N\rightarrow V+2$
transitions, is in the XY universality class.  In order for this
transition to occur, in mean-field theory $\tilde{r}_{T_{6,7}}$
must pass through zero before any of the other members of the set
${\cal S}_N$.  If we restrict the interaction energies to
$f_{pQ}$, $f_{QT}$ and $f_{pQT}$, $\tilde{r}_{T_{6,7}}^{(N)}$ will
never be the smallest in the set. However, higher-order terms of
the form $f_{Q^2T^2}$ can favor the formation of $T_6-T_7$ order
over the others and make this transition possible, with
\begin{eqnarray}
T_6\propto T_7&\sim& |\tilde{r}_{T_{6,7}}^{(N)}-
\tilde{r}_{T_{6,7}c}^{(N)}|^{\beta_{XY}},
\end{eqnarray}
for $\tilde{r}_{T_{6,7}}^{(N)}<\tilde{r}_{T_{6,7}c}^{(N)}$. Because
the of $T_6-T_7$ order drives no other order, the $N_T$ phase is
characterized by nonvanishing $S$, and $\vec{T}_{6,7}$ order.

\subsection{Transitions from the $V$ phase}

In the $V$ phase, three order parameters, $p_3$, $S$, and $T_1$ are
nonzero. Harmonic fluctuations of the other order parameters in this
phase are described by the free energy density
\begin{equation}
{\tilde f}^{(V)} = {\tilde f}_{B_{1,2},T_{6,7}}^{(V)} +
{\tilde f}_{T_{2,3}}^{(V)} + {\tilde f}^{(V)}_{p_{1,2},Q_{3,4},T_{4,5}},
\end{equation}
where
\begin{eqnarray}
{\tilde f}_{B_{1,2}, T_{6,7}}^{(V)} & = & \case{1}{2}
\tilde{r}_{B_{1,2}}^{(V)}(B_1^2 + B_2^2) +
\case{1}{2}\tilde{r}_{T_{6,7}}^{(V)}(T_6^2 + T_7^2)\nonumber \\
&&+ \tilde{\alpha}_{B_1,T_6}^{(V)}(B_1 T_6 + B_2 T_7) \nonumber\\
{\tilde f}_{T_{2,3}}^{(V)} & = &
\case{1}{2}\tilde{r}_{T_{2,3}}^{(V)}(T_2^2 + T_3^2) \\
{\tilde f}^{(V)}_{p_{1,2},Q_{3,4},T_{4,5}}&=&
\case{1}{2}\tilde{r}_{p_{1,2}}^{(V)} (p_1^2 + p_2^2) +
\case{1}{2}\tilde{r}_{Q_{3,4}}^{(V)} (Q_3^2 + Q_4^2)\nonumber\\
&&\hspace{-1.5cm} +\case{1}{2} \tilde{r}_{T_{4,5}}^{(V)}(T_4^2 + T_5^2)
+\tilde{\alpha}_{p_{1,2},Q_{3,4}}^{(V)}(p_1 Q_3 + p_2 Q_4)\nonumber\\
&&\hspace{-1.5cm}+\tilde{\alpha}_{p_{1,2},T_{4,5}}^{(V)}(p_1 T_4 + p_2 T_5)
+\tilde{\alpha}_{Q_{3,4},T_{4,5}}^{(V)}(Q_3 T_4 + Q_4 T_5)\nonumber
\end{eqnarray}
where
\begin{eqnarray}
\tilde{r}_{B_{1,2}}^{(V)} & = &
\tilde{r}_{B_{1,2}}^{(N)}-\case{2}{5}(2w_1+w_3)T_1^2,\nonumber\\
\tilde{r}_{T_{6,7}}^{(V)} & = &
\tilde{r}_{T_{6,7}}^{(N)} + (4u_T - \case{2}{5}v_T)T_1^2,\nonumber\\
\tilde{\alpha}_{B_{1,2},T_{6,7}}^{(V)}& = &\case{4}{\sqrt{15}}w_{QT} T_1 -
\sqrt{\case{2}{3}} w_{pQT} p_3.\nonumber\\
\tilde{r}_{T_{2,3}}^{(V)} & = &
\tilde{r}_{T_{2,3}}^{(N)} + (4u_T + \case{6}{5}v_T)T_1^2,\nonumber\\
\tilde{r}_{p_{1,2}}^{(V)} & = &\tilde{r}_{p_{1,2}}^{(N)}+4u_p p_3^2
+\case{6}{\sqrt{10}}w_pT p_3T_1,\nonumber\\
\tilde{r}_{Q_{3,4}}^{(V)} & = &r_Q - w_Q S + \case{8}{3}u_Q S^2 \nonumber\\
&&-\case{2}{10}(4w_1+2w_2-w_3)T_1^2,\nonumber\\
\tilde{r}_{T_{4,5}}^{(V)} & = &\tilde{r}_{T_{4,5}}^{(N)}
+(4 u_T + \case{6}{25}v_T) T_1^2,\nonumber\\
\tilde{\alpha}_{p_{1,2},Q_{3,4}}^{(V)} &=& -\sqrt{2}w_{pQ}p_3
+\case{1}{\sqrt{5}}w_{pQT}T_1,\nonumber\\
\tilde{\alpha}_{p_{1,2},T_{4,5}}^{(V)}
&=&\tilde{\alpha}_{p_{1,2},T_{4,5}}^{(N)}-2\sqrt{\case{3}{5}}w_{pT} p_3^2,
\nonumber\\
\tilde{\alpha}_{Q_{3,4},T_{4,5}}^{(V)}
&=&-\case{2}{5\sqrt{3}}w_{QT}T_1-2\sqrt{\case{2}{15}}\nonumber\\
&&-\case{2}{15\sqrt{3}}(w_1+6w_2+4w_3)S T_1,
\label{ralphaVs}
\end{eqnarray}
As Fig.\ \ref{flowchart} indicates, there are symmetry-lowering
transitions from the $V$ phase to the $V+2$, $V+3$, and $N+V$
phases. Which one occurs is determined by which the set $S_V =
\{{\tilde r}^{(V)}_{T_{2,3}}, \Delta^{(V)}_{B_{1,2}, T_{6,7}},
\Delta^{(V)}_{p_{1,2},Q_{3,4},T_{4,5}}\}$, where
\begin{mathletters}
\begin{eqnarray}
\Delta_{B_{1,2},T_{6,7}}^{(V)} & = &\tilde{r}_{B_{1,2}}^{(V)}
\tilde{r}_{T_{6,7}}^{(V)} -
\left(\tilde{\alpha}_{B_{1,2},T_{6,7}}^{(V)}\right)^2,
\end{eqnarray}
\end{mathletters}
(and $\Delta^{(V)}_{p_{1,2},Q_{3,4},T_{4,5}}$ unenlightenly
complicated) first reaches its critical value. The nature of the
transition out of the $V$ phase will be determined by which of these
three mean-field reduced temperatures first reaches the respective
critical temperature.

\subsubsection{$V\rightarrow V+3$ transition}

\begin{figure}[bth]
\centering
\setlength{\unitlength}{1mm}
\begin{picture}(150,55)(0,0)
\put(-20,-57){\begin{picture}(150,0)(0,0)
\includegraphics{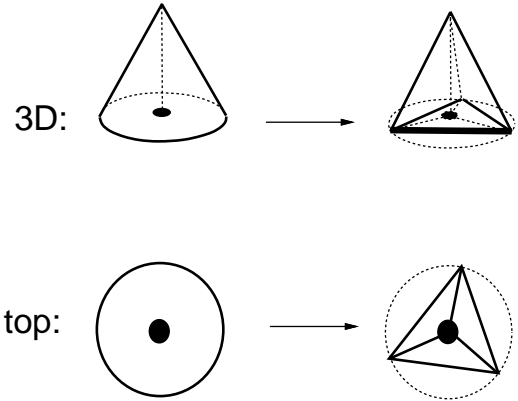}
\end{picture}}
\end{picture}
\caption{A schematic for the $V\rightarrow V+3$ transition.}
\label{V-V3ransition}
\end{figure}

This transition occurs if $\tilde{r}_{T_{2,3}}^{(V)}$ reaches its
critical value $\tilde{r}_{T_{2,3}c}^{(V)}$ before
$\Delta_{B_{1,2},T_{6,7}}^{(V)}$ and
$\Delta^{(V)}_{p_{1,2},Q_{3,4},T_{4,5}}$ reach their respective
critical values, both zero in mean-field theory.  Since there is
rotational invariance in the space defined by $T_2$ and $T_3$, this
transition is in the XY universality class and corresponds to
development of triaxial order ${\vec T}_{2,3}$ in the plane perpendicular to 
the vector order axis $p_3$.  Since no other order is driven by this development
of the ${\vec T}_{2,3}$ order, this order parameter, together with $p_3$, $S$
and $T_1$ (already present in the $V$ phase) completely characterizes
the $V+3$ phase.

\subsubsection{$V\rightarrow V+2$ transition}

\begin{figure}[bth]
\centering
\setlength{\unitlength}{1mm}
\begin{picture}(150,55)(0,0)
\put(-20,-57){\begin{picture}(150,0)(0,0)
\includegraphics{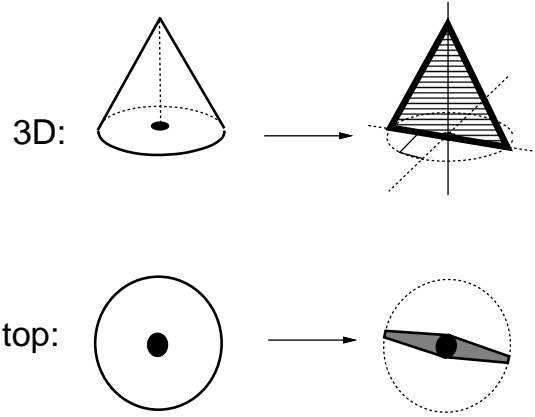}
\end{picture}}
\end{picture}
\caption{A schematic for the $V\rightarrow V+2$ transition.}
\label{V-V2ransition}
\end{figure}

If $\Delta_{B_{1,2},T_{6,7}}^{(V)}$ is the first in the set $S_V$ to
reach its critical, there will be a transition from the $V$ to the
$V+2$ phase signaled by the development of biaxial order in the plane
perpendicular to ${\bf n}$ characterized by a specific linear combination 
of ${\vec B}_{1,2}$ and ${\vec T}_{6,7}$.
Rotational invariance in this plane implies that the transition will
also be in the XY universality class in which an arbitrary linear
combination of $(B_1, T_6)$ and $(B_2, T_7)$ will spontaneously order.

\subsubsection{$V\rightarrow N+V$ transition}

\begin{figure}[bth]
\centering
\setlength{\unitlength}{1mm}
\begin{picture}(150,55)(0,0)
\put(-20,-57){\begin{picture}(150,0)(0,0)
\includegraphics{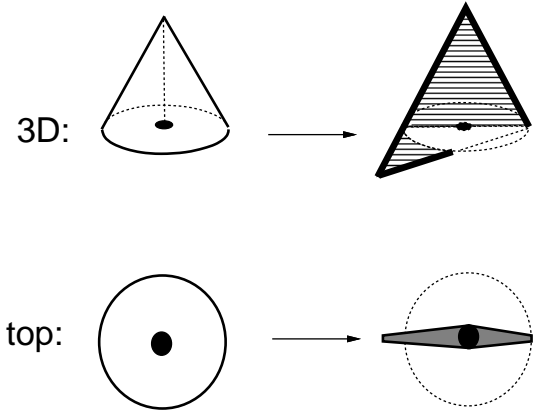}
\end{picture}}
\end{picture}
\caption{A schematic for the $V\rightarrow N+V$ transition.}
\label{V-VNransition}
\end{figure}

If $\Delta^{(V)}_{p_{1,2},Q_{3,4},T_{4,5}}$ is the first in the
set $S_V$ to reach its critical value, there will be a transition
from the $V$ phase to one in which vector (${\vec p}_{1,2}$) and
third-rank tensor (${\vec T}_{4,5}$) order develop in the plane
perpendicular to the already existing vector order, $p_3$. Without
loss of generality we represent this transverse vector ordering by
nonzero $p_1$ and $T_4$. This set of order parameters is invariant
under only one non-trivial operation: reflection from $y$ to $-y$.
Thus the new phase has $C_{1h}$ symmetry and is the $N+V$ phase.
Once these order parameters have been established, they will drive
$B_1$, $T_2$, and $T_6$ nonzero via higher-order terms such as
$f_{pQ}$ and $f_{pT}$. Since the $V \rightarrow N+V$ transition is
controlled by the development of vector order in a plane, we
expect it to be in the $XY$ universality class.

\subsection{Transitions from the $N+2$ phase}

The $N+2$ phase is the standard biaxial nematic phase with
nonvanishing $S$ and an arbitrary linear combination of biaxial $B_1$
and $B_2$ order parameters; without loss of generality, for
convenience, we choose our coordinate system $({\bf m},{\bf l},{\bf
  n})$ so that the biaxial order is described by $B_1$.  As the flow
chart of Fig.\ \ref{flowchart} indicates, the $N+2$ phase can undergo
symmetry-lowering transitions to the $V+2$ and $(N_T + 2)^*$ phases.
Harmonic fluctuations about the $N+2$ phase, which determine the
nature of transitions out of the $N+2$ phase, are described by
\begin{equation}
{\tilde f}^{(N+2)} = {\tilde f}^{(N+2)}_{p_3,T_1,T_6} + {\tilde
f}^{(N+2)}_{p_1,T_2,T_4} + {\tilde f}^{(N+2)}_{p_2,T_3,T_5} + {\tilde
f}^{(N+2)}_{T_7} , \label{harN+2}
\end{equation}
where
\begin{mathletters}
\begin{eqnarray}
{\tilde f}^{(N+2)}_{p_3,T_1,T_6} & = &
\case{1}{2}\tilde{r}_{p_3}^{(N+2)} p_3^2 +
\case{1}{2}\tilde{r}_{T_1}^{(N+2)} T_1^2 +
\case{1}{2}\tilde{r}_{T_6}^{(N+2)} T_6^2\nonumber\\
& &\hspace{-1cm} +\tilde{\alpha}_{T_1,T_6}^{(N+2)} T_1 T_6 +
\tilde{\alpha}_{p_3,T_1}^{(N+2)} p_3 T_1 +
\tilde{\alpha}_{p_3,T_6}^{(N+2)} p_3 T_6,\\
{\tilde f}^{(N+2)}_{p_1,T_2,T_4}& = & \case{1}{2}\tilde{r}_{p_1}^{(N+2)}
p_1^2 + \case{1}{2}\tilde{r}_{T_2}^{(N+2)} T_2^2 +
\case{1}{2}\tilde{r}_{T_4}^{(N+2)} T_4^2\nonumber\\
&&\hspace{-1cm}+\tilde{\alpha}_{T_2,T_4}^{(N+2)} T_2 T_4 +
\tilde{\alpha}_{p_1,T_2}^{(N+2)} p_1 T_2 +
\tilde{\alpha}_{p_1,T_4}^{(N+2)} p_1 T_4,\\ {\tilde
f}^{(N+2)}_{T_7} & = & \case{1}{2} {\tilde r}^{(N+2)}_{T_7} T_7^2 ,
\end{eqnarray}
\label{har2N+2}
\end{mathletters}
where
\begin{eqnarray}
\tilde{r}_{p_3}^{(N+2)} & = & \tilde{r}_{p_3}^{(N)},\nonumber\\
\tilde{r}_{T_1}^{(N+2)} & = &
\tilde{r}_{T_1}^{(N)} - \case{2}{5}(2w_1+w_3)B_1^2 ,\nonumber\\
\tilde{r}_{T_6}^{(N+2)} & = &
\tilde{r}_{T_{6,7}}^{(N)} - \case{2}{3}(2w_1+2w_2+w_3)B_1^2,\nonumber\\
\tilde{\alpha}_{T_1,T_6}^{(N+2)} & = &
\case{4}{\sqrt{15}}w_{QT}B_1-\case{4}{3\sqrt{15}}(2w_1+3w_2-w_3)S B_1,
\nonumber\\
\tilde{\alpha}_{p_3,T_1}^{(N+2)} & = &
\tilde{\alpha}_{p_3,T_1}^{(N)},\nonumber\\
\tilde{\alpha}_{p_3,T_6}^{(N+2)} & = &
\sqrt{\case{2}{3}}w_{pQT} B_1,\nonumber\\
\tilde{r}_{p_1}^{(N+2)} & = &
\tilde{r}_{p_{1,2}}^{(N)} - 2 w_{pQ} B_1,\nonumber\\
\tilde{r}_{T_2}^{(N+2)} & = &
\tilde{r}_{T_{2,3}}^{(N)} - 2(w_1+w_2)B_1^2,\nonumber\\
\tilde{r}_{T_4}^{(N+2)} & = &
\tilde{r}_{T_{4,5}}^{(N)}-\case{4}{5}w_{QT}B_1+\case{2}{15}
\big(w_1(4S-7B_1)\nonumber\\
&&+w_2(4S-B_1)-2w_3(2S+B_1)\big)B_1,\nonumber\\
\tilde{\alpha}_{T_2,T_4}^{(N+2)} & = &
\case{2}{\sqrt{15}}w_{QT}B_1-\case{2}{3\sqrt{15}}
\big(2w_1 S\nonumber\\
&&+3w_2(2S-B_1)+w_3(2S-3B_1)B_1\big),\nonumber\\
\tilde{\alpha}_{p_1,T_2}^{(N+2)} & = & -w_{pQT} B_1,\nonumber\\
\tilde{\alpha}_{p_1,T_4}^{(N+2)} & = &
\tilde{\alpha}_{p_{1,2},T_{4,5}}^{(N)}+\case{1}{\sqrt{15}}w_{pQT} B_1,
\nonumber\\
\tilde{r}_{T_7}^{(N+2)} & = & \tilde{r}_{T_{6,7}}^{(N)} -
\case{2}{3}(2w_1 - w_3)B_1^2,\label{rN+2s}
\end{eqnarray}
and ${\tilde f}^{(N+2)}_{p_2,T_3,T_5}$ is obtained from ${\tilde
  f}^{(N+2)}_{p_1,T_2,T_4}$ by replacements: $p_1\rightarrow p_2$,
$T_2\rightarrow T_3$, $T_4\rightarrow T_5$, $B_1\rightarrow -B_1$.

Following our earlier analysis of other transitions, we introduce the
determinants $\Delta_{p_3,T_1,T_6}^{(N+2)}$,
$\Delta_{p_1,T_2,T_4}^{(N+2)}$, and $\Delta_{p_2,T_3,T_5}^{(N+2)}$ of
the matrices, respectively defined by the coefficients in the free
energy densities ${\tilde f}^{(N+2)}_{p_3,T_1,T_6}$, ${\tilde
f}^{(N+2)}_{p_1,T_2,T_4}$ and ${\tilde f}^{(N+2)}_{p_2,T_3,T_5}$.
Which transition from the $N+2$ phase takes place is determined by the
set ${\cal S}_{N+2} =
\{\tilde{r}_{T_7}^{(N+2)},
\Delta_{p_3,T_1,T_6}^{(N+2)}, \Delta_{p_1,T_2,T_4}^{(N+2)},
\Delta_{p_2,T_3,T_5}^{(N+2)}\}$.

\subsubsection{$N+2 \rightarrow V+2$ transition}

\begin{figure}[bth]
\centering
\setlength{\unitlength}{1mm}
\begin{picture}(150,55)(0,0)
\put(-20,-57){\begin{picture}(150,0)(0,0)
\includegraphics{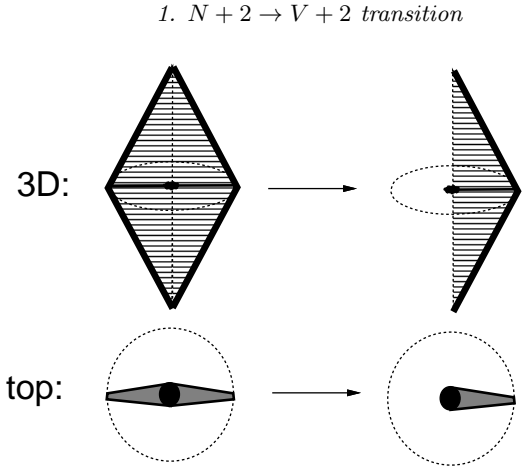}
\end{picture}}
\end{picture}
\caption{A schematic for the $N+2\rightarrow V+2$ transition.}
\label{N2-V2ransition}
\end{figure}

Interestingly, there are three routes from the $N+2$ phase to a phase
with $V+2$ symmetry.  In the first route, the nematic axis along ${\bf
  n}$, which in the $N+2$ phase is invariant under ${\bf n}
\rightarrow - {\bf n}$, is converted to a vector axis with the
development of $p_3$, $T_1$, and $T_6$ order. The biaxial order of the
$N+2$ phase persists, resulting in the $V+2$ phase with vector order
along the two-fold axis ${\bf n}$. In mean-field theory, this
transition to $V+2$ phase takes place when
$\Delta_{p_3,T_1,T_6}^{(N+2)}$ is the first in the set ${\cal
  S}_{N+2}$ to pass through zero.

In the second route, vector order develops along the direction
${\bf m}$, corresponding to the maximum eigenvalue of the nematic
order parameter $Q^{ij}$ in the plane perpendicular to ${\bf n}$.
In this route, which, in mean-field theory occurs when
$\Delta_{p_1,T_2,T_4}^{(N+2)}$ is the first in ${\cal S}_{N+2}$ to
go through zero, the $p_1$, $T_2$, and $T_4$ order parameters
become nonzero.

In the third route, $\Delta_{p_2,T_3,T_5}^{(N+2)}$ is the first in
${\cal S}_{N+2}$ to go through zero, and vector order develops
along the ${\bf l}$ axis perpendicular to ${\bf m}$, defined by
the eigenvector of $Q^{ij}$ with minimum eigenvalue in the plane
perpendicular to ${\bf n}$. It is the $p_2$, $T_3$, and $T_5$
order parameters that become nonzero at the transition.

In all three versions of the $N+2\rightarrow V+2$ transition the
vector ${\bf p}$ order develops along one of the three two-fold
symmetry axes defined by the biaxial order of the $N+2$ phase, the
eigenvectors of $Q^{ij}$.  In each case, ${\bf p}$ can develop
either a positive or negative value along an axis already chosen
by the $N+2$ phase.  Consequently, these $N+2\rightarrow V+2$
transitions are in the Ising universality class, with well-known
critical properties.

\subsubsection{$N+2 \rightarrow (N_T + 2)^*$ transition}
\label{N2toXi}

\begin{figure}[bth]
\centering
\setlength{\unitlength}{1mm}
\begin{picture}(150,55)(0,0)
\put(-20,-57){\begin{picture}(150,0)(0,0)
\includegraphics{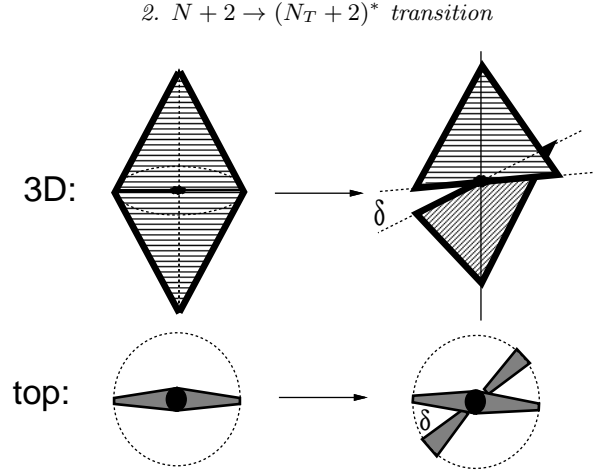}
\end{picture}}
\end{picture}
\caption{A schematic for the $N+2\rightarrow (N_T+2)^*$
transition.} \label{N2-N_T2transition}
\end{figure}

For our choice of the ${\bf m}$-${\bf l}$ axes, for which the biaxial order of
the $N+2$ phase is characterized by $B_1\neq0$ and $B_2=0$, another
symmetry-reducing transition from the $N+2$ phase is signaled by the
appearance of the $T_7$ order. It corresponds to the $N+2\rightarrow
(N_T+2)^*$ and takes place when the reduced temperature
$\tilde{r}_{T_7}^{(N+2)}$ is the smallest in the set ${\cal S}_{N+2}$. For a
different choice of axes, in which $B_2\neq0$ and $B_1=0$ characterize biaxial
order, this transition would instead correspond to development of $T_6$ order.
And, more generally, in an $N+2$ phase characterized by a particular linear
combination of $B_1$ and $B_2$, i.e., by ${\vec B}_{1,2}$ a corresponding {\em
orthogonal} linear combination of $T_6$ and $T_7$, i.e., the ${\vec
T}_{6,7}$ order parameter, such that ${\vec B}_{1,2}\cdot {\vec T}_{6,7}=0$ 
develops at the $N+2\rightarrow (N_T+2)^*$ transition.  

To understand the nature of this transition it is helpful to first
visualize the $N+2$ phase as a collection of orientationally
ordered but positionally disordered {\em planar} diamond units,
each consisting of two, leg-to-leg banana molecules, illustrated
in Figs.\ \ref{N_Tphases}(b),\ \ref{tetrafig}(c), and
\ref{N2-N_T2transition}. One can choose the $\{{\bf n, m, l}\}$
triad such that the diamonds lie in the ${\bf n}$-${\bf m}$ plane
and, therefore, that their biaxiality is characterized by
nonvanishing $S$ and $B_1$ order parameters. Now the transition to
the $(N_T+2)^*$ phase corresponds to {\em counter}, out-of-plane
twisting about the ${\bf n}$ axis of the two diamond-forming
banana molecules. It is signaled by the development of the $T_7$
order parameter, with the twist angle $\delta$ in the range 
$0\leq\delta\leq\pi/2$ given by
\begin{equation}
\tan\delta=T_7/B_1.\label{twist_angle}
\end{equation}
We note that the lower limit of this range $\delta=0$ corresponds
to the planar diamonds of the achiral $N+2$ phase and the upper
limit, $\delta=\pi/2$ (reached only in the limit
$T_7\rightarrow\infty$ or $B_1 = 0$) corresponds to the $N_T$
phase, which is also achiral, and lacks biaxial order . In
contrast all other values of the twist angle $0\le\delta\le\pi/2$,
corresponding to a nonzero value of $T_7$ describe a {\em
spontaneously} induced chirality of the $(N_T + 2)^*$ phase.  Since
the sign of the twist angle (and correspondingly the sign of
$T_7$) can be either positive (right-handed) or negative
(left-handed), in mean-field theory the $N+2\rightarrow (N_T + 2)^*$ 
transition is clearly in the well-known Ising universality class (but see
below).

Finally, we observe that in a phase which spontaneously breaks
chiral (mirror) symmetry, a tensor, totally antisymmetric in all
its indices (akin to the well-known $\epsilon_{ijk}$ tensor) must
spontaneously emerge. It is easy to verify that in the chiral 
$(N_T + 2)^*$ phase, most generally characterized by finite $S$, $B_1$,
$B_2$, $T_6$ and $T_7$ order parameters, $\Xi_{ijk}$ is given by
\begin{mathletters}
\begin{eqnarray}
\Xi_{ijk} &=& Q^{is} B^{jt} T^{stk} + Q^{js} B^{kt} T^{sti} +
Q^{ks} B^{it} T^{stj},\\
&=& \case{1}{\sqrt{6}}\,S (B_1 T_7-B_2 T_6)\,\epsilon_{ijk},
\end{eqnarray}
\end{mathletters}
is indeed such a fully {\em antisymmetric} tensor, which develops
spontaneously from our theory, that is based solely on fully {\em
symmetric} tensors $Q^{ij}$ and $T^{ijk}$.

The existence of a fully anti-symmetric third-rank tensor allows
for invariant couplings linear in spatial derivative.  For
example, a term of the form
\begin{equation}
\Xi_{ijk} Q^{il} \nabla_j Q^{kl} \sim T_7 \nv \cdot \gradv \times
\nv
\label{chiralXi}
\end{equation}
is permitted, where the right hand side represents the dominant
part near the transition where $S$ and $B_1$ are effectively
constant, $T_7$ is small, and $\nv$ is the Frank director.  The
elastic energy of the $N+2$ phase includes the usual twist energy
$K_2 [\nv \cdot \gradv \times \nv]^2$.  There are other terms in
the elastic free energy arising from the biaxial order of the
$N+2$ phase, and they may influence the nature of the ground state
of the $(N_T+2)^*$ phase.  The existence of terms such as that of
Eq.\ \ref{chiralXi} that are linear in spatial gradients implies
that the ground state of the $(N_T+2)^*$ phase will be spatially
inhomogeneous.  The simplest chiral inhomogeneous phase we can
imagine is a cholesteric $(N_T+2)^*$ phase in which ${\bf n}$
rotates in a helical fashion as in the standard cholesteric with
pitch $P$ and pitch wave number $q_0= 2\pi/P$ which near the
transition to the $N+2$ phase at temperature $T_c$ scales as
\begin{equation}
q_0 \sim T_7/K_2 \sim |\Delta T|^{1/2}
\end{equation}
in mean-field theory where $\Delta T = T- T_c$. In a true critical
theory, $q_0$ will also scale to zero as a power of $|\Delta T|$
provided the transition remains second order. In the presence of
fluctuations, the $T_7 \nv \cdot
\gradv \times \nv$ term is likely to modify the universality class
of this transition to something other than the naively expected
Ising universality class (possibly even driving the transition
first-order), but we have not, however, analyzed the critical
theory in detail. If molecules themselves are chiral, or a chiral 
dopant is added, then the
$N+2$ phase will be a chiral $(N+2)^*$ phase with a nonvanishing
$q_0$. Chirality will act like an external ordering field (like
the magnetic field of an Ising model) for $q_0$ with strength $h$,
and if the $N+2$-to-$(N_T+2)^*$ transition is second order, one
can expect $q_0(T,h)$ to scale as
\begin{equation}
q_0(T,h) = |T-T_c|^{\beta_q}f(h/|T-T_c|^{\Delta}),
\end{equation}
where $T_c$ is the transition temperature, $\Delta$ is the
external-field gap exponent, and $\beta_q$ is a critical exponent,
which in mean-field theory is the order-parameter exponent $\beta
= 1/2$.

\subsection{Transitions from the $N+3$ phase}

As we indicate in Fig.\ \ref{flowchart}, the $N+3$ phase,
described by coexistence of the uniaxial order parameter $S$ and
the transverse (to the nematic axis) triaxial order parameter
${\vec T}_{2,3}\equiv (T_2 , T_3)$, can undergo symmetry-lowering
transitions to the $V+2$, $V+3$, and $N+V$ phases.

These transitions are all signaled by the development of vector
order ${\bf p}$. They are distinguished by whether this vector
order is {\em along} ($N+3\rightarrow V+3$ transition), {\em
transverse} to ($N+3\rightarrow V+2$ ($\equiv N+1$) transition),
or at an arbitrary polar angle $0\le\delta\le\pi/2$ to the
uniaxial ${\bf n}$ axis, singled out in the nematic phase.
Although it is convenient to think of these transitions as driven
by the vector order ${\bf p}$, more precisely they are driven by
specific linear combinations of the vector and other order
parameter, linearly coupled to ${\bf p}$.

It is convenient to choose the ${\bf m}$-${\bf l}$ axes so that
the triaxial order of the $N+3$ phase, in the plane perpendicular
to ${\bf n}$ phase is described by $T_2\neq0$ and $T_3=0$. For
this choice, the harmonic fluctuations about the $N+3$ phase are
described by
\begin{equation}
{\tilde f}^{(N+3)} = {\tilde f}^{(N+3)}_{p_3,T_1} + {\tilde
f}^{(N+3)}_{p_1,B_1,T_4} + {\tilde f}^{(N+3)}_{Q_{3,4},T_{6,7}},
\label{fN+3total}
\end{equation}
where
\begin{eqnarray}
{\tilde f}^{(N+3)}_{p_3, T_1} & = &
\case{1}{2}\tilde{r}_{p_3}^{(N+3)} p_3^2 +
\case{1}{2}\tilde{r}_{T_1}^{(N+3)} T_1^2 +
\tilde{\alpha}_{p_3,T_1}^{(N+3)} p_3 T_1,\nonumber\\
{\tilde f}^{(N+3)}_{p_{1,2},B_{1,2},T_{4,5}} & = &
\case{1}{2}\tilde{r}_{p_{1,2}}^{(N+3)}(p_1^2 + p_2^2) +
\case{1}{2}\tilde{r}_{B_{1,2}}^{(N+3)}(B_1^2 + B_2^2)\nonumber\\
&&\hspace{-1.5 cm} +\case{1}{2}\tilde{r}_{T_{4,5}}^{(N+3)}(T_4^2 +
T_5^2)
+\tilde{\alpha}_{p_{1,2},B_{1,2}}^{(N+3)}(p_1 B_1 - p_2 B_2 )\nonumber\\
&&\hspace{-1.5cm}+\tilde{\alpha}_{p_{1,2},T_{4,5}}^{(N+3)}(p_1 T_4
+ p_2 T_5) +\tilde{\alpha}_{B_{1,2},T_{4,5}}^{(N+3)}(B_1 T_4 - B_2
T_5),\nonumber\\
\hspace{-0.3cm}{\tilde f}^{(N+3)}_{Q_{3,4},T_{6,7}} & = &
\case{1}{2}\tilde{r}_{Q_{3,4}}^{(N+3)}(Q_3^2 + Q_4^2)
+\case{1}{2}\tilde{r}_{T_{6,7}}^{(N+3)}(T_6^2 + T_7^2)\nonumber\\
&&+\tilde{\alpha}_{Q_{3,4},T_{6,7}}^{(N+3)}(Q_3T_6 - Q_4T_7)
\label{fN+3},
\end{eqnarray}
with the coefficients
\begin{eqnarray}
\tilde{r}_{p_3}^{(N+3)} & = & \tilde{r}_{p_3}^{(N)},\nonumber\\
\tilde{r}_{T_1}^{(N+3)} & = &
\tilde{r}_{T_1}^{(N)} + (4u_T+\case{6}{5}v_T)T_2^2 ,\nonumber\\
\tilde{\alpha}_{p_3,T_1}^{(N+3)} & = &
\tilde{\alpha}_{p_3,T_1}^{(N)},\nonumber\\
\tilde{r}_{p_{1,2}}^{(N+3)} & = &
\tilde{r}_{p_{1,2}}^{(N)},\nonumber\\
\tilde{r}_{B_{1,2}}^{(N+3)} & = &
\tilde{r}_{B_{1,2}}^{(N)}-2(w_1+w_2)T_2^2,\nonumber\\
\tilde{r}_{T_{4,5}}^{(N+3)} & = &
\tilde{r}_{T_{4,5}}^{(N)} + (4u_T+\case{4}{5}v_T)T_2^2,\nonumber\\
\tilde{\alpha}_{p_{1,2},B_{1,2}}^{(N+3)} & = &
-w_{pQT} T_2,\nonumber\\
\tilde{\alpha}_{p_{1,2},T_{4,5}}^{(N+3)} & = &
\tilde{\alpha}_{p_{1,2},T_{4,5}}^{(N)}\nonumber\\
\tilde{\alpha}_{B_{1,2},T_{4,5}}^{(N+3)}& = &
\case{2}{\sqrt{15}}w_{QT} T_2 -
{\case{4}{3\sqrt{15}}}(w_1+3w_2+w_3)S T_2,\nonumber\\
\tilde{r}_{Q_{3,4}}^{(N+3)} & = & r_Q - w_Q S + \case{8}{3}u_Q S^2
-\case{1}{2}w_1 T_2^2,\nonumber\\
\tilde{r}_{T_{6,7}}^{(N+3)} & = &
\tilde{r}_{T_{6,7}}^{(N)} + 4u_T T_2^2,\nonumber\\
\tilde{\alpha}_{Q_{3,4},T_{6,7}}^{(N+3)}& = &
-\case{1}{\sqrt{3}}w_{QT} T_2 -{\case{2}{3\sqrt{3}}}(6w_1-w_3)S
T_2.\label{rN+3s}
\end{eqnarray}
Transitions out of the $N+3$ phase are controlled by the set
${\cal
S}_{N+3}=\{\Delta_{p_3,T_1}^{(N+3)},\Delta_{p_{1,2},B_{1,2},
T_{4,5}}^{(N+3)},\Delta_{Q_{3,4},T_{6,7}}^{(N+3)}\}$ of
determinants of the harmonic coefficients that can be read off
from ${\tilde f}^{(N+3)}$ above.

\subsubsection{$N+3 \rightarrow V+3$ transition}

\begin{figure}[bth]
\centering
\setlength{\unitlength}{1mm}
\begin{picture}(150,55)(0,0)
\put(-20,-57){\begin{picture}(150,0)(0,0)
\includegraphics{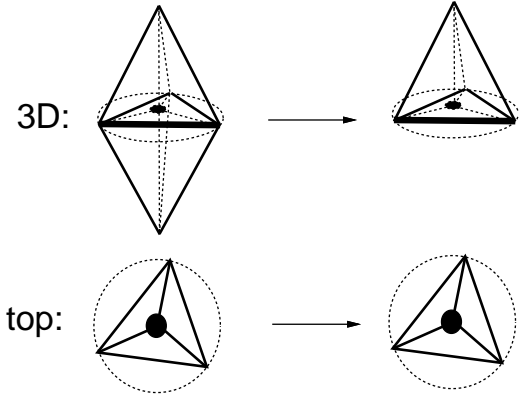}
\end{picture}}
\end{picture}
\caption{A schematic for the $N+3\rightarrow V+3$ transition.}
\label{N3-V3transition}
\end{figure}

The $N+3 \rightarrow V+3$ transition occurs when
$\Delta_{p_1,T_3}$ is smallest in the set ${\cal S}_{N+3}$. In
this transition, vector order develops along the nematic ${\bf n}$
axis to produce a linear combination of $p_3$ and $T_1$ order,
with the latter explicitly induced as a third harmonic of the
$p_3$ order. The discrete, ${\cal Z}_2$, ${\bf n} \rightarrow
-{\bf n}$ symmetry, characterizing the $N+3$ phase is lost at this
transition. Consequently, the $N+3 \rightarrow V+3$ transition is
in the familiar Ising universality class.

\subsubsection{$N+3 \rightarrow V+2$ ($=N+1$) transition}

\begin{figure}[bth]
\centering
\setlength{\unitlength}{1mm}
\begin{picture}(150,55)(0,0)
\put(-20,-57){\begin{picture}(150,0)(0,0)
\includegraphics{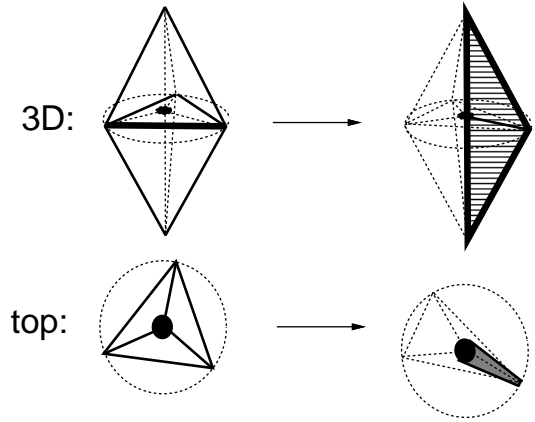}
\end{picture}}
\end{picture}
\caption{A schematic for the $N+3\rightarrow N+1$ transition.}
\label{N3-N1transition}
\end{figure}

$V+2$ order develops from $N+3$ order by spontaneously favoring
one of the three equivalent directions in the plane perpendicular
to ${\bf n}$ and, thereby, converting transverse {\em triaxial}
order of $N+3$ into transverse {\em vector }\, order of $N+1$,
equivalently described by $V+2$. With ${\bf p}$ chosen to be along
${\bf m}$ , the transverse vector order is signaled by the
development of finite $p_1$ order parameter. As can be seen from
${\tilde f}^{(N+3)}_{p_{1,2},B_{1,2},T_{4,5}}$, Eq.\ (\ref{fN+3}),
order parameters $B_1$ and $T_4$, associated with $p_1$ and
linearly coupled to it, are also simultaneously induced at the
$N+3 \rightarrow V+2$ transition. For a more generic choice of the
orientation of ${\bf m}$-${\bf l}$ within the plane perpendicular
to ${\bf n}$, the $V+2$ phase is described by a specific linear
combination of $\vec{p}_{1,2}\equiv (p_1,p_2)$, 
$\vec{B}_{1,2}\equiv (B_1,B_2)$, and $\vec{T}_{4,5}\equiv
(T_4,T_5)$, obtained from the basic set $\{p_1,B_1,T_4\}$ by a
planar rotation about the nematic axis ${\bf n}$.

The $N+3 \rightarrow V+2$ transition is the exit of choice out of
the $N+3$ phase, when $\Delta_{p_{1,2},B_{1,2},T_{4,5}}^{(N+3)}$
is the smallest in the ${\cal S}_{N+3}$ set. Since the $N+3$ phase
is characterized by three equivalent directions (in the plane
perpendicular to the nematic axis ${\bf n}$) along which vector
order ${\bf p}$ can develop, we expect this transition to be in
the universality class of the $3$-states Potts model, believed to
be weakly first-order in three dimensions and continuous in two.

These symmetry based expectations are born out by our detailed
computations, which show that the rotational degeneracy in the ${\bf
  m}$-${\bf l}$ plane, which is present in the harmonic free energy,
${\tilde f}^{(N+3)}_{p_{1,2},B_{1,2},T_{4,5}}$, Eq.\ (\ref{fN+3})
is lifted by energy contributions of the
form %
\begin{eqnarray}
\delta{\tilde f}^{(N+3)}_{p_{1,2},B_{1,2},T_{4,5}}&=&
-\alpha_p\cos3\phi-\alpha_B\cos6\phi,
\label{cos3phi}
\end{eqnarray}
where $\phi$ is the angle between the developing transverse vector
order parameters ${\vec p}_{1,2}$ (as well as ${\vec B}_{1,2}$ 
and ${\vec T}_{4,5}$)
and the ${\bf m}$ axis defined by triaxial order of the $N+3$ phase,
\begin{mathletters}
\begin{eqnarray}
\alpha_p &=& \left(\case{1}{2}w_{pT}p_{1,2}^3 -
\case{4}{5\sqrt{15}}v_T T_{4,5}^3\right) T_2\ ,\label{alpha_p}\\
\alpha_B &=& w_{Q^3T^2} B_{1,2}^3 T_2^2\ ,\label{alpha_B}
\label{alpha_pB}
\end{eqnarray}
\end{mathletters}
and $\delta{\tilde f}^{(N+3)}_{p_{1,2},B_{1,2},T_{4,5}}$ arises from
the following nonlinear couplings
\begin{mathletters}
\begin{eqnarray}
f_{pT} &=& - w_{pT}\ p^i p^j p^k T^{ijk}\label{fp3T}\\
f_{Q^3T^2} &=& - w_{Q^3T^2}\ Q^{i_1 j_1} Q^{i_2 j_2} Q^{i_3 j_3}
T^{i_1 i_2 i_3} T^{j_1 j_2 j_3}\label{fQ3T2}\\
f_{T^4} &=& v_T\ T^{i_1 i_2 i_3} T^{i_1 i_4 i_5}
T^{i_2 i_4 i_6} T^{i_3 i_5 i_6}.\label{fvT}
\end{eqnarray}
\end{mathletters}
The three degenerate minima of the free energy $\delta{\tilde
f}^{(N+3)}_{p_{1,2},B_{1,2},T_{4,5}}$, Eq.\ (\ref{cos3phi}), for
the vector order parameter to settle into, correspond precisely to
the three equivalent states of the $3$-state Potts model,
supporting our expectation that the $N+3\rightarrow V+2$
transition is in the $3$-state Potts model universality class.

\subsubsection{$N+3 \rightarrow N+V$ transition}
\label{N+3toN+V}

\begin{figure}[bth]
\centering
\setlength{\unitlength}{1mm}
\begin{picture}(150,55)(0,0)
\put(-20,-57){\begin{picture}(150,0)(0,0)
\includegraphics{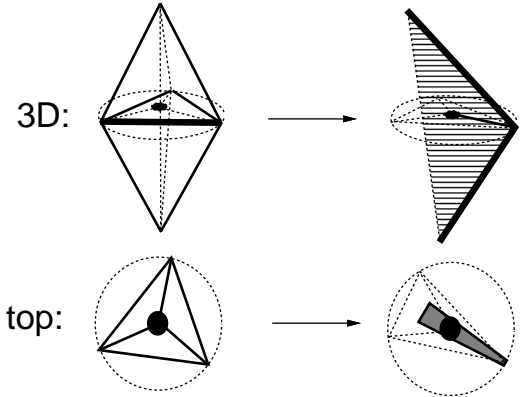}
\end{picture}}
\end{picture}
\caption{A schematic for the $N+3\rightarrow N+V$ transition.}
\label{N3-NVtransition}
\end{figure}

The transition that occurs when ${\tilde\Delta}_{Q_{3,4},T_{6,7}}^{(N+3)}$ is
the smallest of the set ${\cal S}_{N+3}$ is, as we shall see, to the $N+V$
phase, but the development of all of the order parameters characterizing this
phase is complicated. At the {\em harmonic} level in ${\vec T}_{6,7}$ and
${\vec Q}_{3,4}$, ${\tilde f}^{(N+3)}$ possesses a rotational invariance in
the plane perpendicular to the nematic axis ${\bf n}$. However, because the
$N+3$ phase is characterized by 3-fold order in the $xy$ plane perpendicular
to ${\bf n}$ (aligned along the $z$ axis), this continuous symmetry will be
reduced to a discrete clock-symmetry. 

It is convenient to focus on ordering of $Q_3$ and $T_6$.  The established
$N+3$ order can be viewed as an equilateral triangle that favors alignment of
${\bf m}$ towards one of its vertices.  Thus one might predict the symmetry of
a 3-state clock model.  However, the apolar (nematic along $z$ axis) nature of
the $N+3$ state ensures that the free energy ${\tilde f}^{(N+3)}(Q_3,T_6)$ is
invariant under ${\bf n}\rightarrow - {\bf n}$.  This, combined with
transformation properties of $Q_3$ and $T_6$ under ${\bf n} \rightarrow -{\bf
n}$, with $Q_3, T_6\rightarrow -Q_3, -T_6$ gaurantees an additional Ising
symmetry of the free energy ${\tilde f}^{(N+3)}(Q_3,T_6)={\tilde
f}^{(N+3)}(-Q_3,-T_6)$.  
This then leads to a total of $6$ free-energetically degenerate states. 
Integrating out the ${\vec T}_{6,7}$ order parameter and focusing on ${\vec
Q}_{3,4}$ alone, shows that the $6$ states correspond to 
alignment of ${\bf m}_Q$ along $3$ vertices and $3$ edge-bisectors of the 
equilateral triangle defined by the $N+3$ phase.  Thus there is a six-state
clock symmetry described by a coupling proportional to $T_2^2 Q_3^6$, and we
expect the $N+3 \rightarrow N+V$ transition to be in the universality class of
the six-state clock model, which apart from irrelevant variables is in the
universality class of the $XY$-model.  Once $Q_3$ and $T_6$ order is
established, $T_4$, $p_1$, and $B_1$ order is driven by couplings of the form
$Q_3^3 B_1$ and others.  Thus, $p_1 \sim B_1 \sim T_4 \sim Q_3^2$.  Then in
mean-field theory, $B_1$ order will drive $p_3$ and $T_1$ order via couplings
of the form $B_1 T_6 p_3$.  Thus $p_3 \sim T_1 \sim T_6^3 \sim Q_3^3$.  

\subsection{Transitions from the $T$ phase}

The Tetrahedratic $T$ phase with $T_d$ symmetry is characterized by a
nonvanishing arbitrary linear combination of $T_6$ and $T_7$, which we
collectively refer to as $\vec{T}_{6,7}$.  In this section, without
lost of generality, we will choose the orientation of the ${\bf
  m}-{\bf l}$ axes so that $T_7$ is the only nonvanishing order
parameter in the $T$ phase.  The harmonic free energy density for
fluctuations from the $T$ phase can be expressed as
\begin{equation}
{\tilde f}^{(T)} = {\tilde f}_{Q_1}^{(T)} + {\tilde f}_{Q_2}^{(T)} + {\tilde
f}_{p_{1,2,3}Q_{3,4,5}}^{(T)} ,
\end{equation}
where
\begin{eqnarray}
{\tilde f}_{Q_1}^{(T)} & = & \case{1}{2}\tilde{r}_{Q_1}^{(T)} Q_1^2, \nonumber\\
{\tilde f}_{Q_2}^{(T)} & = & \case{1}{2}\tilde{r}_{Q_2}^{(T)} Q_2^2, \label{fT2}\\
{\tilde f}_{p_{1,2,3}Q_{3,4,5}}^{(T)} & = &
\case{1}{2}\tilde{r}_{Q_{3,4,5}}(Q_3^2+Q_4^2+Q_5^2)+
\case{1}{2} r_p(p_1^2+p_2^2+p_3^2)\nonumber\\
& & - \case{1}{\sqrt{3}}w_{pQT} T_7 (p_1 Q_4 + p_2 Q_3 + p_3 Q_5 )
\nonumber\label{f(T)pQ}
\end{eqnarray}

with the coefficients given by
\begin{mathletters}
\begin{eqnarray}
\tilde{r}_{Q_1}^{(T)}&=& r_Q - \case{1}{3}(2w_1-w_3) T_7^2,\\
\tilde{r}_{Q_2}^{(T)}&=& r_Q - \case{1}{3}(2w_1-w_3) T_7^2,\\
\tilde{r}_{Q_{3,4,5}}&=& r_Q - \case{1}{3}(2w_1 + 2w_2 + w_3)
T_7^2 .
\end{eqnarray}
\label{rTs}
\end{mathletters}
The structure of ${\tilde f}^{(T)}$ implies that there are three
symmetry-lowering transitions from the $T$ phase, driven respectively
by fluctuations in $Q_1$, $Q_2$, and a linear combination of pairs
$p_1$ and $Q_4$, $p_2$ and $Q_3$, and $p_3$ and $Q_5$.  Which of these
transitions occurs is determined by the set ${\cal S}_T =
\{\tilde{r}_{Q_1}^{(T)},\ \tilde{r}_{Q_2}^{(T)},
\ \Delta_{p_{1,2,3}Q_{3,4,5}}^{(T)}\}$.

In the presence of $T_7$ order, all of the $p^i_\mu$ and
$Q^{ij}_\mu$ order parameters exhibit third-order invariants.
Since they play an important role in determining the nature of
transitions from the $T$ phase, we display them here:
\begin{eqnarray}
f_{Q^3} & = & w_Q\big[\sqrt{\case{3}{2}} Q_1 (Q_2^2 + Q_5^2)
- \case{\sqrt{3}}{2\sqrt{2}}Q_1 (Q_3^2 + Q_4^2)\nonumber\\
& & - \case{3}{2\sqrt{2}}Q_2 (Q_3^2 - Q_4^2)
-\case{1}{\sqrt{6}} Q_1^3 - \case{3}{\sqrt{2}}Q_3 Q_4 Q_5\big]
\label{fQ3}\nonumber\\
f_{p^3 T} & = & - \sqrt{6} w_{pT} T_7 p_1 p_2 p_3\label{fp3}.
\end{eqnarray}
We note the appearance of the $Q_3 Q_4 Q_5$ term in $f_{Q^3}$, which
can be paired with the $p_1 p_2 p_3$ product in $f_{p^3 T}$.  There
are of course also third order terms in $Q$, whose coefficients are
proportional to powers of $T_7$, which, for small $T_7$, are
subdominant to $w_Q$ term that we displayed above. Because these
higher order terms do not qualitatively change our results, with their
effects accounted for by an effective $w_Q$ coupling, we will not
consider them here.

\subsubsection{$T \rightarrow N_T$ transition}

\begin{figure}[bth]
\centering
\setlength{\unitlength}{1mm}
\begin{picture}(150,35)(0,0)
\put(-20,-57){\begin{picture}(150,0)(0,0)
\includegraphics{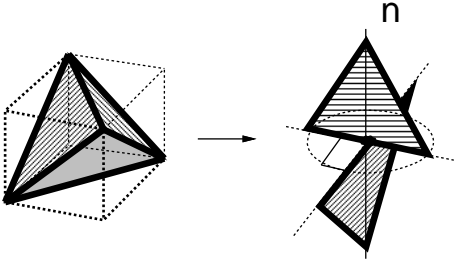}
\end{picture}}
\end{picture}
\caption{A schematic for the $T\rightarrow N_T$ transition.}
\label{T-N_Ttransition}
\end{figure}

As illustrated in Fig.\ \ref{flowchart}, one of the three possible
symmetry-reducing transitions out of the Tetrahedratic $T$ phase
is the $T\rightarrow N_T$ phase transition. The $ N_T$ order
develops by favoring one of the $3$ two-fold axes of the
tetrahedron through the growth of uniaxial $Q^{ij}$ order along
that axis as shown in Figs.~\ref{T-N_Tphases},
\ref{tetrafig}, and \ref{T-N_Ttransition}.
In our parameterization, we focus on the two-fold axes defines by
$\pm {\bf n}$. In this case, the $T\rightarrow N_T$ transition is
signaled by the development of $Q_1$ (or equivalently $S$) order.
Because of the existence of the third-order $Q_1^3$ invariant,
Eq.\ (\ref{fQ3}), this transition is generically first-order. It
occurs in mean-field theory when $\tilde{r}_{Q_1}^{(T)}(T_7)= -
\tilde{w}_Q^2(T_7)/12\tilde{u}_Q(T_7)$, where the third- and
fourth-order couplings $\tilde{w}_Q(T_7)$ and $\tilde{u}_Q(T_7)$
can in principal depend on strength of the $T$ order,
characterized by $T_7$ order parameter. Because no other order
parameters is explicitly induced at this transition, and with the
convenient choice of the ${\bf m}$-${\bf l}$ axes (that we have
made here) the resulting $ N_T$ phase is fully characterized by
the $T_7$ and $S$ order parameters.

\subsubsection{$T \rightarrow (N_T + 2)^*$ transition}

\begin{figure}[bth]
\centering
\setlength{\unitlength}{1mm}
\begin{picture}(150,35)(0,0)
\put(-20,-57){\begin{picture}(150,0)(0,0)
\includegraphics{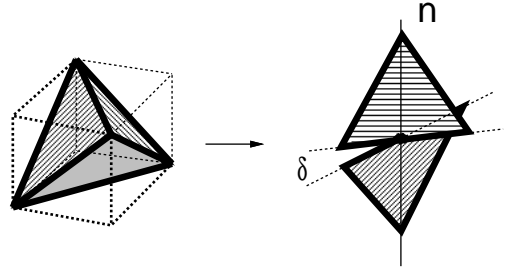}
\end{picture}}
\end{picture}
\caption{A schematic for the $T\rightarrow (N_T+2)^*$ transition.}
\label{T-N_T2transition}
\end{figure}

If, for a $T$ phase characterized by the $T_7$ order parameter,
$Q_2$ (or equivalently $B_1$) orders before $Q_1$ or the $p_1-Q_3$
combination, the transition from the $T$ phase is to the
spontaneously chiral $(N_T + 2)^*$ phase. If we focus on the ${\bf
m}$-${\bf l}$ plane, $T_7$ order displays biaxiality $(m^i l^j +
m^j l^i)$, with principle axes at $\pi/4$ relative to the
biaxiality of the $B_1$ order characterized by $(m^i m^j - l^i
l^j)$. Since $T_7$ and $B_1$ are different order parameters, a
simple rotation to define a new biaxial axis is impossible, and
the result is that the reflection invariance in the ${\bf
m}$-${\bf l}$ plane is spontaneously lost as $B_1$ develops. The
lack of reflection symmetry is the characteristic feature of the
chiral $(N_T + 2)^*$ phase, discussed in Sec.\ \ref{N2toXi}. In
analogy to the development of biaxial order from the Isotropic
phase, once $B_1$ orders, the $Q_1 Q_2^2$ coupling in $f_{Q^3}$
drives the development of $S$ (or $Q_1$) order. Thus, the $(N_T +
2)^*$ phase has nonvanishing $S$, $B_1$, and $T_7$ (or
equivalently $S$, $B_2$, and $T_6$) order.  Although the
$T\rightarrow (N_T + 2)^*$ transition is continuous within
mean-field theory, based on the experience with the development of
uni- and bi-axial orders from the Isotropic phase, we expect that
here too, the $T\rightarrow N_T$ transition will always preempt
the $T\rightarrow (N_T + 2)^*$ transition.

Within second-order mean-field theory, $B_1 \sim |\Delta
T|^{1/2}$, $S \sim B_1^2 \sim |\Delta T|$, and $\Xi_{ijk} \sim T_7
B_1 S \sim |\Delta T|^{3/2}$.  Since the $T$ phase has elastic
energies resisting spatial variations of the $\nv$, $\mv$, and
$\lv$ directions, the wavenumber of the cholesteric structure of
the cholesteric $(N_T+2)^*$ phase will scale as $q_0 \sim |\Delta
T|^{3/2}$ in mean-field theory.

\subsubsection{$T\rightarrow V+3$ transition}

\begin{figure}[bth]
\centering
\setlength{\unitlength}{1mm}
\begin{picture}(150,35)(0,0)
\put(-20,-57){\begin{picture}(150,0)(0,0)
\includegraphics{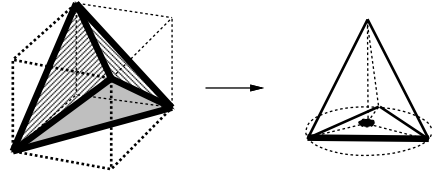}
\end{picture}}
\end{picture}
\caption{A schematic for the $T\rightarrow V+3$ transition.}
\label{T-V3transition}
\end{figure}

The ordering of linear combinations of $(p_1,Q_4)$, $(p_2,Q_3)$,
and $(p_3, Q_5)$, which we will refer to as $p-Q$ ordering, leads
to the $V+3$ phase. This can be seen by observing that the third
order potential $f_{p^3,T}$ in Eq.\ (\ref{fp3}) favors $p_1 = \pm
p_2 =\pm p_3$ with relative signs determined by the sign of
$w_{pT}T_7$. Examination of Eq.\ (\ref{fQ3}) show that similar
considerations apply to $Q_3$, $Q_4$ and $Q_5$. Thus, the vector
${\bf p}$ will align along the $(1,1,1)$ or a symmetry equivalent
axes, i.e., along a three-fold axis of the tetrahedron
characterizing the $T$ phase. Thus, to discuss the phase
transition signaled by the onset of $p-Q$ order, its is useful to
transform to a new coordinate system with ${\bf n}'$ along the
$(1,1,1)$ axis, which is most easily achieved by a rotation about
the $(1,-1,0)$ axis. Under this rotation we find that our basis
transforms according to
\begin{mathletters}
\begin{eqnarray}
{\bf m} &=& \case{1}{2}(1+\case{1}{\sqrt{3}}){\bf m}'-
\case{1}{2}(1-\case{1}{\sqrt{3}}){\bf l}' -
\case{1}{\sqrt{3}}{\bf n}',\\
{\bf l} &=& -\case{1}{2}(1-\case{1}{\sqrt{3}}){\bf m}'+
\case{1}{2}(1+\case{1}{\sqrt{3}}){\bf l}' -
\case{1}{\sqrt{3}}{\bf n}',\\
{\bf n} &=& \case{1}{\sqrt{3}}({\bf m}'+{\bf l}'+{\bf n}').
\end{eqnarray}
\end{mathletters}
Straightforward algebraic manipulations then yield
\begin{equation}
I_7^{ijk} = \case{\sqrt{5}}{3}{I'}_1^{ijk}-
\case{\sqrt{2}}{3}({I'}_2^{ijk} + {I'}_3^{ijk}),
\end{equation}
so that in the rotated coordinate system,
\begin{mathletters}
\begin{eqnarray}
T'_1 &=& \case{\sqrt{5}}{3} T_7,\\
T'_2 &=& -\case{\sqrt{2}}{3} T_7,\\
T'_3 &=& -\case{\sqrt{2}}{3} T_7.
\end{eqnarray}
\end{mathletters}
In this rotated basis the polar order is described by ${\bf
  p}=p_3'{\bf n'}$, also inducing the nematic order $S'$ through the
$f_{pQ}$ coupling.  Thus, the phase produced by such $p-Q$ ordering
from the $T$ phase has nonvanishing $p_3'$, $S'$, $T_1'$, $T_2'$, and
$T_3'$ order, which describes polar order along the ${\bf n}'$ axis
and triaxial order in the ${\bf m}$-${\bf l}$ plane perpendicular to
it; a rotation by $\pi/12$ about the $(1,1,1)$ axis can be used to
remove $T_3'$ component of the triaxial order. Thus, the phase
produced by $p-Q$ ordering inside the Tetrahedratic $T$ phase is
indeed the previously discussed $V+3$ phase.

\subsection{Transitions from the $ N_T$ phase}

The $ N_T$ phase is characterized by nonvanishing nematic order
parameter, $S$, and an arbitrary linear combination of the
tetrahedratic order parameters, $T_6$ and $T_7$, which we collectively
call $\vec{T}_{6,7}$.  These define the directions of the orthonormal
triad $({\bf m}, {\bf l}, {\bf n} )$, which we can for convenience
always pick to have $T_6$ vanish, with the tetrahedratic order
completely characterized by the value of $T_7$.

The free energy ${\tilde f}^{(N_T)}$ describing {\em harmonic}
fluctuations in the $N_T$ phase can be expressed as a sum of three
independent parts:
\begin{equation}
{\tilde f}^{(N_T)} = {\tilde f}_{B_1}^{(N_T)} + {\tilde
f}_{p_3,B_2,T_1}^{(N_T)}+
{\tilde f}_{p_{1,2},Q_{3,4},T_{2,3},T_{4,5}}^{(N_T)},
\end{equation}
where
\begin{eqnarray}
{\tilde f}_{B_1}^{(N_T)} &=&
\case{1}{2}\tilde{r}_{B_1}^{(N_T)} B_1^2,\nonumber\\
{\tilde f}_{p_3,T_1,B_2}^{(N_T)} & = &
\case{1}{2} \tilde{r}_{p_3}^{(N)} p_3^2 +
\case{1}{2}\tilde{r}_{B_2}^{(N_T)} B_2^2 +
\case{1}{2} \tilde{r}_{T_1}^{(N_T)} T_1^2\nonumber\\
&&\hspace{-1.5cm}
+\tilde{\alpha}_{p_3,B_2}^{(N_T)} p_3 B_2 +
\tilde{\alpha}_{p_1,T_3}^{(N)} p_3 T_1+
\tilde{\alpha}_{B_2,T_1}^{(N_T)} B_2 T_1,\nonumber\\
{\tilde f}_{p_{1,2},Q_{3,4},T_{2,3},T_{4,5}}^{(N_T)}& = &
\case{1}{2}\tilde{r}_{p_{1,2}}^{(N)} (p_1^2 + p_2^2) +
\case{1}{2} \tilde{r}_{Q_{3,4}}^{(N_T)}(Q_3^2 + Q_4^2)\nonumber\\
&&\hspace{-3cm}
+\case{1}{2}\tilde{r}_{T_{2,3}}^{(N_T)} (T_2^2 + T_3^2)
+\case{1}{2} \tilde{r}_{T_{4,5}}^{(N_T)}(T_4^2 + T_5^2)\\
&&\hspace{-3cm}
+\tilde{\alpha}_{p_{1,2},Q_{3,4}}^{(N_T)}(p_1 Q_4 + p_2 Q_3)
+\tilde{\alpha}_{p_{1,2},T_{4,5}}^{(N)}(p_1 T_4 + p_2 T_5)\nonumber\\
&&\hspace{-3cm}
+\tilde{\alpha}_{B'_2,T'_1}^{(N_T)}
\big[Q_3(\case{1}{\sqrt{3}} T_3-\case{1}{\sqrt{5}} T_5)
+Q_4(\case{1}{\sqrt{3}} T_2-\case{1}{\sqrt{5}} T_4)\big],\nonumber\\
&&\hspace{-3cm} +\tilde{\alpha}_{Q_{3,4},T_{4,5}}^{(N_T)} (Q_3
T_5+Q_4 T_4) +\tilde{\alpha}_{T_{2,3},T_{4,5}}^{(N_T)} (T_2
T_4+T_3 T_5), \nonumber
\end{eqnarray}
with
\begin{eqnarray}
\tilde{r}_{B_{1}}^{(N_T)} & = & \tilde{r}_{B_{1,2}}^{(N)}
-\case{2}{3} (2w_1 - w_3)T_7^2,\nonumber\\
\tilde{r}_{B_{2}}^{(N_T)} & = & \tilde{r}_{B_{1,2}}^{(N)}
-\case{2}{3} (2w_1 + 2w_2 + w_3)T_7^2,\nonumber\\
\tilde{r}_{T_1}^{(N_T)} & = & \tilde{r}_{T_1}^{(N)} +
(4u_T-\case{2}{5}v_T)T_7^2,\nonumber\\
\tilde{\alpha}_{p_3,B_2}^{(N_T)}&=&
-\sqrt{\case{2}{3}}w_{pQT}T_7,\nonumber\\
\tilde{\alpha}_{B_2,T_1}^{(N_T)}&=&
\case{4}{\sqrt{15}}w_{QT}T_7 - \case{4}{3\sqrt{15}}(2w_1+3w_2-w_3)S T_7,\nonumber\\
\tilde{r}_{Q_{3,4}}^{(N_T)}&=& 2r_Q+\case{8}{3} u_Q S^2
-\case{1}{3}(2w_1+2w_2+w_3)T_7^2,\nonumber\\
\tilde{r}_{T_{2,3}}^{(N_T)} & = & \tilde{r}_{T_{2,3}}^{(N)}
+4u_T T_7^2,\label{rT23Sigma}\nonumber\\
\tilde{r}_{T_{4,5}}^{(N_T)} & = & \tilde{r}_{T_{4,5}}^{(N)}
+(4u_T+\case{4}{15}v_T)T_7^2,\label{rT45Sigma}\nonumber\\
\tilde{\alpha}_{p_{1,2},Q_{3,4}}^{(N_T)} & = &
-\case{1}{\sqrt{3}}w_{pQT}T_7,\nonumber\\
\tilde{\alpha}_{B'_2,T'_1}^{(N_T)}&=&
w_{QT}T_7 + \case{1}{3}(w_1-2w_3)S T_7,\nonumber\\
\tilde{\alpha}_{Q_{3,4},T_{4,5}}^{(N_T)}&=&
-\case{4}{3\sqrt{5}} w_2 S T_7,\nonumber\\
\tilde{\alpha}_{T_{2,3},T_{4,5}}^{(N_T)}&=& \case{2}{\sqrt{15}}
v_T T_7^2 .\label{rs3}
\end{eqnarray}

As can be seen from the structure of ${\tilde f}^{(N_T)}$ above,
there are three possible symmetry-reducing transitions that can
take place out of the $N_T$ phase. In contrast to the nematic
phase with $D_{\infty h}$ symmetry in which fluctuations in the
biaxial fields $B_1$ and $B_2$ are degenerate, the $N_T$ phase
with nonvanishing ${\vec T}_{6,7}$ order breaks the degeneracy of
fluctuations in $B_1$ and $B_2$. For our choice of ${\bf m}$-${\bf
l}$ axes, with $T_6=0$, the tetrahedratic order parameter $T_7$
couples the biaxial order parameter $B_2$ to the vector order
along ${\bf n}$, described by $p_3$ and its third harmonic $T_1$.
Hence two (of the three) transitions are the ordering of $B_1$,
and the ordering of a linear combination of $p_3$, $B_2$ and
$T_1$. A third possible transition out of the $N_T$ phase is the
development of vector order {\em transverse} to nematic axis ${\bf
n}$ described by an arbitrary linear combination of $p_1$ and
$p_2$. In the $N_T$ phase with nonzero $T_7$, harmonic
fluctuations in $p_1$ are coupled to those of $Q_4$ and $T_4$, and
harmonic fluctuations in $p_2$ are coupled to those of $Q_3$ and
$T_5$.  As a result, the development of transverse vector order
from the $N_T$ phase is accompanied by a specific linear
combination of these higher-order order parameters.

Which of these three transitions takes place first is determined
by minimum determinant in the set ${\cal
S}_{N_T}=\{\Delta_{B_1},\Delta_{p_3,B_2,T_1}^{(N_T)},
\Delta_{p_{1,2},Q_{3,4},T_{2,3},T_{4,5}}^{(N_T)}\}$ of
determinants of the harmonic coefficients, that can be read off
from ${\tilde f}^{(N_T)}$ above.

\subsubsection{$N_T \rightarrow (N_T + 2)^*$ transition}

\begin{figure}[bth]
\centering
\setlength{\unitlength}{1mm}
\begin{picture}(150,55)(0,0)
\put(-20,-57){\begin{picture}(150,0)(0,0)
\includegraphics{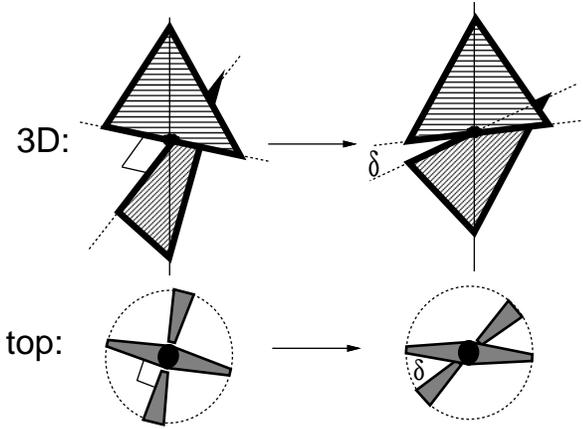}
\end{picture}}
\end{picture}
\caption{A schematic for the $N_T\rightarrow (N_T+2)^*$
transition.} \label{N_T-N_T2transition}
\end{figure}

Development of biaxial order, $B_1$, in the presence of $T_7$
(with $T_6=0$) and $S$ order is quite similar to the development
of $T_7$ order in the presence of $B_1$ order, discussed in Sec.\
\ref{N2toXi}, and corresponds to spontaneous chiral symmetry
breaking of the $N_T \rightarrow (N_T + 2)^*$ transition. As can
be seen from the structure of ${\tilde f}^{(N_T)}$, no other order
parameters are induced at this transition and, as discussed above,
the resulting $(N_T + 2)^*$ phase is characterized by nonzero $S$,
$B_1$ and $T_7$ order parameters. Because it is the underlying
$B_1\rightarrow -B_1$, ${\cal Z}_2$ symmetry of the $N_T$ phase,
that is broken when $B_1$ orders, the $N_T \rightarrow (N_T +
2)^*$ transition is in the Ising universality class if the linear
gradient coupling proportional to $\Xi_{ijk}$ is ignored.  The
latter term, aside from potentially modifying the critical properties of
this transition, leads to a pitch wavenumber in the cholesteric
$(N_T+2)^*$ phase that scales in mean-field theory as $q_0 \sim
B_1 \sim |\Delta T |^{1/2}$ just below the $N_T \rightarrow (N_T
+2)^*$ transition.

\subsubsection{$N_T \rightarrow V+2$ transition}

\begin{figure}[bth]
\centering
\setlength{\unitlength}{1mm}
\begin{picture}(150,55)(0,0)
\put(-20,-57){\begin{picture}(150,0)(0,0)
\includegraphics{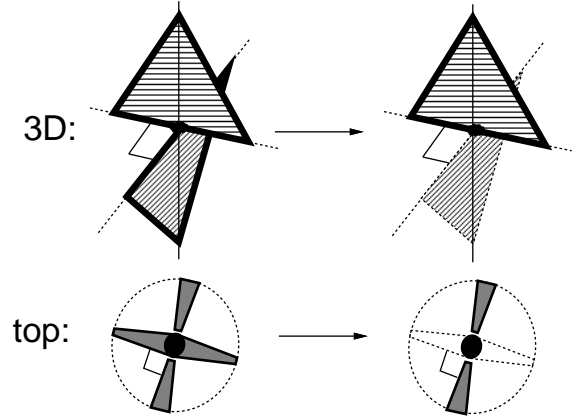}
\end{picture}}
\end{picture}
\caption{A schematic for the $N_T\rightarrow V+2$ transition.}
\label{N_T-V2transition}
\end{figure}

The $N_T \rightarrow V+2$ transition takes place via the
development of vector order ${\bf p}=p_3{\bf n}$ along the
nematically ordered ${\bf n}$ axis.  Since in the presence of
$T_7$ and $S$ such longitudinal vector order is coupled to $B_2$
and $T_1$, the $N_T \rightarrow V+2$ is also accompanied by the
development of $B_2$ and the $T_1$ order parameter. Because the
free energy of the $N_T$ phase is invariant under ${\bf
n}\rightarrow -{\bf n}$, the development of $p_3$ longitudinal
order can be of either sign and the transition is in the
Ising universality class.

\subsubsection{$N_T \rightarrow N+V$ transition}

A third possible transition out of the $N_T$ phase takes place
when {\em transverse} vector order $\vec{p}_{1,2}$ develops. As
can be seen from the form of ${\tilde
f}_{p_{1,2},Q_{3,4},T_{2,3},T_{4,5}}^{(N_T)}$, the development of
$\vec{p}_{1,2}$ is accompanied by biaxial order $\vec{Q}_{3,4}$
and a linear combination of $\vec{T}_{2,3}$ and $\vec{T}_{4,5}$
order parameters. The resulting phase has a $C_{1h}$ symmetry, and
we therefore identify it with the previously discussed $N+V$
phase. At the harmonic level, the free energy appears to be $O(2)$
invariant, with respect to rotation of $\vec{p}_{1,2}$. However,
in the presence of $T_7$, nonlinearities in $\vec{p}_{1,2}$, break
this rotational invariance. A lowest order such symmetry breaking
nonlinearity is given by
\begin{eqnarray}
f_{\rm symm. breaking}^{N_T}&=& Q^{i_1,i_2} T^{i_1,j_1,k_1}
T^{i_2,j_2,k_2} p^{j_1} p^{k_1} p^{j_2} p^{k_2},\nonumber\\
&=&\case{4}{9}S T_7^2 p_1^2 p_2^2,
\end{eqnarray}
It introduces into the $O(2)$ symmetry harmonic free energy of
$\vec{p}_{1,2}$ a well-studied cubic symmetry breaking
anisotropy. Based on these studies\cite{Rudnick}, we therefore expect
the $N_T\rightarrow N+V$ transition to be driven first-order by thermal
fluctuations.\cite{commentN+V*}

\subsection{Transitions from the $V+2$ phase}

As can be seen from the flow-chart, Fig.\ \ref{flowchart}, the
biaxial vector $V+2$ phase, characterized for example by finite
$p_3$, $S$, $B_1$, $T_1$, $T_6$ order parameters and by $C_{2v}$
symmetry, can further lower its symmetry in two ways.  It can
undergo a transition to the $N+V$ phase via the development of
polar order, characterized by ${\vec p}_{1,2}$, ${\vec Q}_{3,4}$, and 
${\vec T}_{4,5}$, order parameters 
along one of the biaxial axis perpendicular to
the $V$ ($p_3$) order.  Alternatively, it can undergo a transition to
the $(V_T +2)^*$ phase via the development of $T_7$
order. Both the $V+2 \rightarrow N+V$ and the $V+2 \rightarrow
(V_T + 2)^*$ are expected to be in the Ising universality class
because in both cases it is ${\cal Z}_2$ symmetry that being
broken.

\subsection{$V+3 \rightarrow N+V$ transition}

The three-fold symmetry in the plane transverse to the vector ($p_3$)
axis of the $V+3$ phase can be spontaneously broken with e.g., biaxial
order in this plane driving the transition and other parameters
(listed in Table 1) also condensing. We therefore expect the $V+3
\rightarrow N+V$ transition to be in the three-states Pott's model
universality class.

\subsection{$(N_T + 2)^* \rightarrow (V_T + 2)^*$  transition}

\begin{figure}[bth]
\centering
\setlength{\unitlength}{1mm}
\begin{picture}(150,55)(0,0)
\put(-20,-57){\begin{picture}(150,0)(0,0)
\includegraphics{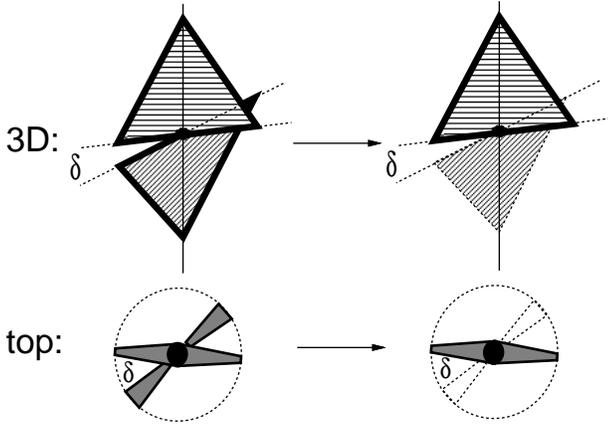}
\end{picture}}
\end{picture}
\caption{A schematic for the $(N_T+2)^*\rightarrow (V_T+2)^*$
transition.} \label{N_T2-V_T2transition}
\end{figure}

The final transition that we will comment on is polar ordering
transition from the nonpolar $N_T + 2^*$ phase. Because of the
present nematic order in $N_T + 2^*$ state, polar order breaks
${\cal Z}_2$ symmetry and we expect $(N_T + 2)^* \rightarrow (V_T +
2)^*$ transition to be in the Ising universality class.

\section{New Smectic Phases}
\label{smectic}

Our primary interest in this paper has been in fluid (spatially homogeneous)
but anisotropic liquid-crystalline phases of bent-core molecules, particularly
in phases with exotic symmetries not encountered in systems of rod- or
plate-like mesogens.  Our work, however, suggests the possibility of smectic
or layered analogs of these exotic fluid phases.  In these putative phases,
which we will explore in more detail in a separate publication\cite{future}
the smectic layer normal ${\bf N}$ provides an additional direction that may
or may not coincide with a symmetry direction of the fluid phase. Most of the
banana smectic phases classified to date are based upon $V+2$ order embedded
in smectic layers either in identical configurations (as in the Sm$C_S P_F$
phase in the notation\cite{B2phases} of ref.\ \cite{Link}) neighboring layers
or in alternating configurations (as in the Sm$C_AP_A$ phase) in neighboring
layers. 
\begin{figure}[bth]
  \centering \setlength{\unitlength}{1mm}
\begin{picture}(150,150)(0,0)
\put(-16,-10){\begin{picture}(150,0)(0,0)
\includegraphics{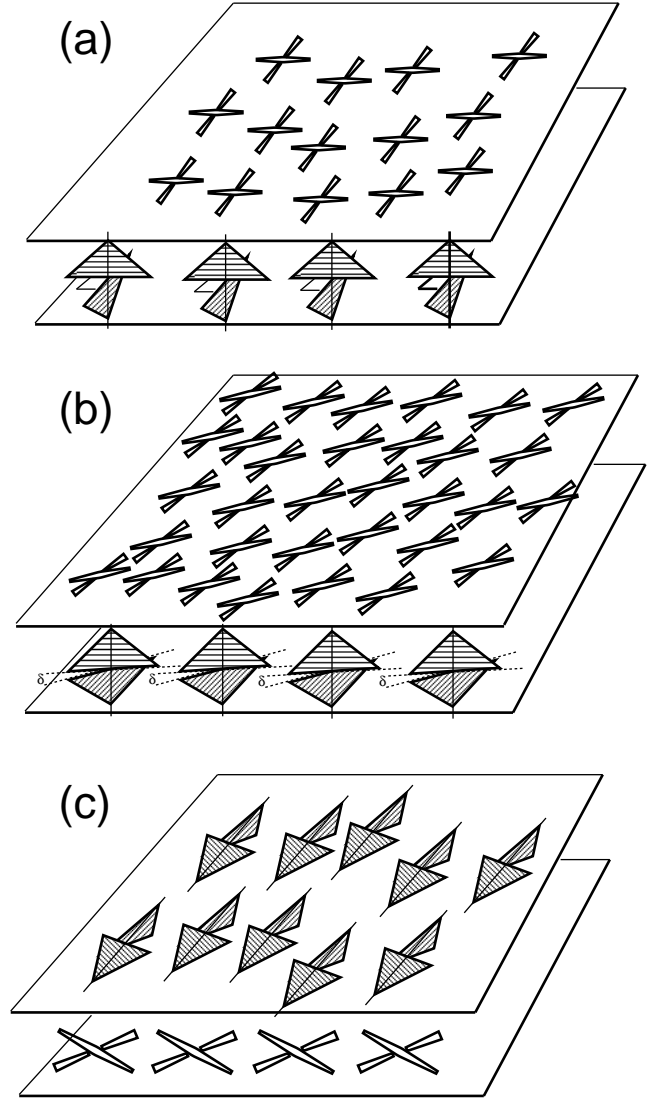}
\end{picture}}
\end{picture}
\caption{Schematic of smectic phases (a) Sm$A_{N_T}$, (b) Sm$A^*_{||}(N_T+2)$,
  (c) SmA$^*_\perp(N_T+2)$.}
\label{SmA_Tphases}
\end{figure}
Each layer is then characterized by the molecular directions ${\bf n}$, ${\bf
m}$, and ${\bf l}$ and by ${\bf N}$.  Though the $V+2$ fluid phase is not
chiral, each layer of smectic phases derived from it can be chiral if, for
example, ${\bf m}$ lies in the smectic layer and ${\bf n}$ is tilted relative
to ${\bf N}$ as is the case in the globally chiral Sm$C_A P_A$ and Sm$C_S P_F$
phases, and the Sm$C_S P_A$ and Sm$C_A P_F$ phases with chirality alternative
in adjacent layers\cite{B2phases}.  

New types of smectics can arise from layering of $N_T$,
$(N_T+2)^*$, and $(V_T+2)^*$ fluid phases.  In the simplest of
these phases, which we label Sm$A_{N_T}$ and depict one layer of
in Fig.\ \ref{SmA_Tphases}(a), each layer has $N_T$ symmetry with
${\bf n}$ parallel to ${\bf N}$ and ${\bf m}$ along a common
direction in each layer.  This phase has $D_{2d}$ point group
symmetry.  Many variants of it are imaginable.  For example, the
${\bf m}$-axis could rotate by $\pi/2$ from layer to layer, or the
${\bf n}$ axis could tilt relative to ${\bf N}$ to produce a
Sm$C^*(N_T)$ phase, which like the Sm$C^*$ phase would be chiral.
Layered phases formed from the $(N_T +2)^*$ fluid phase are
chiral.  In one such phase, the Sm$A_{||}^*(N_T+2)$ phase, a layer
of which is depicted in Fig.\ \ref{SmA_Tphases}(b), the ${\bf n}$
axis is parallel to ${\bf N}$, and the biaxial ${\bf m}$ axis
rotates from layer to layer like the ${\bf c}$ director in the
Sm$C^*$ phase. If the pitch of the twist structure is very long,
this phase would appear to be a biaxial smectic.  The recently
identified biaxial Sm$A$ phase\cite{Madhusudana} in mixtures of
banana and rod-like mesogens may correspond to the very long-pitch
Sm$A_{||}^*(N_T+2)$ phase. An alternative version of a smectic
$(N_T+2)^*$ phase, the Sm$A_{\perp}^*(N_T+2)$ phase has ${\bf n}$
in the plane of the layers as depicted for a single layer in Fig.\
\ref{SmA_Tphases}(c).  Since this phase is chiral, ${\bf n}$ will
rotate in a helical fashion from layer to layer.  A more complex
$(N_T+2)$ smectic-C like phase, with ${\bf n}$ making an angle 
other than $0$ or $\pi/2$ with respect to ${\bf N}$ are also possible.

\section{Summary and Conclusions}
\label{Conclusions}

In this paper, we have presented a comprehensive study of liquid
phases of achiral bent-core liquid-crystal molecules. Using
symmetry we enumerated all possible orientationally-ordered liquid
phases, classified them by subgroups of the rotation group O(3)
under which they are invariant, and constructed Landau mean-field
theory describing these phases and transitions between them. One
primary conclusion of our work is that in addition to the vector
($p^i$) and second-rank nematic ($Q^{ij}$) order parameters, a
third-rank tensor order parameter $T^{ijk}$, representing
third-mass moment, is necessary to characterize novel
liquid-crystal phases of banana molecules, such as for example the
$N_T$ phase with $D_{2d}$ symmetry, the tetrahedratic $T$
phase, and the spontaneously chiral nematic $(N_T + 2)^*$ and its
chiral polar analog $(V_T+2)^*$. In these phases the chiral symmetry
is spontaneously broken by ``condensation'' of the biaxial
$\vec{B}_{1,2}$ and tetrahedratic $\vec{T}_{6,7}$ order parameters
with a nonvanishing angle $0 < \delta < \pi/2$ between their
respective principle axes ${\bf m}-{\bf l}$ (see
Fig.\ref{N_Tphases}(b)).

The $N_T$ phase is neither uniaxial nor biaxial but instead
exhibits an invariance with respect to a four-fold improper
rotation consisting of a rotation through $\pi/2$ about the
$z$-axis followed by an reflection $z \rightarrow -z$.  The $T$
phase is invariant under the $T_d$ symmetry group of a
tetrahedron.  Like the isotropic phase, its second rank dielectric
tensor is isotropic, making it optically isotropic, but unlike the
isotropic phase, it has a nonvanishing second order nonlinear
susceptibility $\chi^{(2)}_{ijk} \sim T^{ijk}$ such that there is
a second order contribution to the polarization $P^{(2)}_i
=\chi^{(2)}_{ijk} E_j E_k$ where $E_i$ is the electric field.

The chiral nematic $(N_T +2)^*$ phase, like the traditional chiral
cholesteric and blue phases, will exhibit periodic spatial
modulations of the direction of molecular alignment.  Unlike the
transition from the isotropic to the cholesteric phase, 
the transitions to the
$(N_T+2)^*$ phase (from the biaxial $N+2$ phase, the tetrahedratic $T$
phase and the $N_T$ phase with $D_{2d}$ symmetry) are second-order, 
at least in mean-field theory.  The pitch of the
cholesteric structure of the $(N_T+2)^*$ phase diverges as these
second-order transitions are approached and thus changes rapidly
with temperature.  Since the $(N_T+2)^*$ phase spontaneously
breaks chiral symmetry, the state initially formed upon cooling
from the higher-symmetry phase will consist of domains of opposite
chirality separated by domain wall that will coarsen over time.
Chiral dopants (or distortions of bent-core mesogens to make them
chiral) render all phases chiral and, in particular, induce a
cholesteric pitch of a particular sign on the chiral extensions
$(N+2)^*$, $T^*$, and $N_T^*$, of the $(N+2)$, $T$, and $N_T$
phases.  Thus chiral dopants act like an external field in an
Ising ferromagnet, favoring a particular sign of chirality (rather than a
particular sign of spin), and the transitions from the $(N+2)^*$,
$T^*$, and $N_T^*$ phases to the $(N_T+2)^*$ phase will be
analogous to the Ising transition in an external magnetic field.
In principle, for sufficiently large chirality, blue phases, with
two orthogonal twist axes can also appear in $(N_T + 2)^*$.  In
addition to these properties of the nonpolar $(N_T + 2)^*$ phase,
the chiral {\em polar} $(V_T+2)^*$ phase will exhibit spontaneous
ferroelectricity, a liquid state that has been a holy-grail in
liquid-crystal research dating back to Louis Pasteur.
Light-scattering, circular dichroism and switching with a weak 
electric field would be natural experimental probes for these
spontaneously chiral states.

These chiral phases are particularly interesting because their smectic
analogs, the chiral Sm$C_A P_A$, Sm$C_S P_A$, Sm$C_A P_F$, and Sm$C_S P_F$
(the four B$_2$ phases)\cite{Link,Walba,B2phases} have been realized in banana
liquid crystal molecules, generating significant excitement in the
ferroelectric liquid-crystal community.  Our work suggests that smectic
positional order is not necessary and that spatially homogeneous spontaneously
chiral liquid-crystal phases are generically possible.  

Although, most\cite{bananaNematics} experimental systems of banana
molecules studied so far, appear to undergo direct first-order
transitions into smectic phases, our work suggests that this
situation does not have to be the case, and a rich phase structure
and hierarchy of continuous transitions studied here is possible.
It is our hope that results presented here will stimulate searches
in experiments and simulations for banana materials that exhibit
novel orientationally-ordered liquid phases predicted here.

\acknowledgments

Leo Radzihovsky was supported by the National Science Foundation
through MRSEC Program at the University of Colorado at Boulder
under \#DMR-9809555, and by the A.P.  Sloan and David and Lucille
Packard Foundations, and acknowledges hospitality of Harvard's
Department of Physics, where this work was completed.  Tom
Lubensky was supported by the NSF through grant DMR00-96531. The
authors thank Noel Clark and Darren Link for numerous discussions.
%


\end{multicols}

\begin{references}
\bibitem{B2phases} In a more faithful nomenclature put forth by the
  Boulder group in Ref.\ \cite{Link}, the so-called B$_2$ phase actually
  refers to four
  different smectic-C phases SmC$_i$P$_j$ of banana molecules. These
  are distinguished by the values of the $i,j$ labels.  The $i$
  label is either $S$ for Synclinic or $A$ for Anticlinic
  indicating, respectively, that molecules in neighboring layers
  tilt in the same or opposite directions.  The $j$ label
  is either $F$ for Ferro or $A$ for Anti-ferro indicating,
  respectively that dipoles in neighboring layers are either
  parallel or anti-parallel. The corresponding
  four phases are: the interlayer synclinic (same tilt from layer to
  layer) and ferroelectric Sm$C_S P _F$ (homochiral) phase, the
  interlayer anticlinic and ferroelectric Sm$C_A P_F$ phase, the
  interlayer synclinic and antiferroelectric Sm$C_S P_A$ phase, and
  the interlayer anticlinic and antiferroelectric Sm$C_A P_A$ phase,
  all of which have now been observed.
\bibitem{mft_exponent} In mean-field theory, a noncritical order
  parameter explicitly induced by spontaneous ordering of another
  order parameter has its growth with reduced temperature and field,
  completely determined by the critical exponents controlling the
  behavior of the critical order parameter. Canonical example is an
  XY-model where ordering of $\psi_1$ explicitly induces higher
  harmonic order parameters, such as e.g., $\psi_n$ through coupling
  $\psi_1^n \psi_n^*The chiral  + c.c.$. However, despite of the irrelevance of
  this coupling for the critical behavior of $\psi_1$ (noncritical
  $\psi_n$ can be simply integrated out, finitely renormalizing bare
  parameters of the $H[\psi_1]$), because of the nonlinear nature of
  this coupling operator, beyond mean-field theory, it, as well as
  susceptibility $\chi_n$ for $\psi_n$ generically receive nontrivial
  fluctuation corrections and therefore have independent and
  nontrivial scaling dimensions. These can be shown to
  lead\cite{Aharony} to breakdown of mean-field $\beta_n= n\beta_1$
  relation, as is most strikingly seen for an XY-model in two
  dimensions.
\bibitem{fel2} In the preceding reference, Fel studied the
  $I\rightarrow T$ transitions and many of the interesting properties
  of the $T$ phase using a Landau theory for a third-rank
  symmetric-traceless tensor order parameter. He failed, however, to
  identify all of the invariants in the Landau theory and, therefore,
  failed to identify the $I \rightarrow N+3$ transition.  He also
  missed the second quartic invariant in $T^{ijk}$ (the $v_T$ term in
  our Eq.\ref{fT}), and therefore incorrectly identified the $I
  \rightarrow T$ transition as being in the $O(7)$ universality class.
  For correct analysis, see Ref.\onlinecite{ITpaper}
\bibitem{B12positive} It is easy to show using Eq.\ (\ref{rs2}) for
  $\tilde{r}_{B_{1,2}}^{(N)}$ that within the quartic $f_Q$ model,
  Eq.\ (\ref{fQ}), valid only close to the $I$-$N$ transition (and even then,
  only if it is weakly first-order, so that $S$ is small), the
  $\tilde{r}_{B_{1,2}}^{(N)}=6w_Q S$ is strictly positive, precluding
  an $N\to N+2$ transition.  However, a more general free energy
  density of $Q^{ij}$ (e.g., including higher order nonlinearities in
  $Q^{ij}$) can be shown to lead to $\tilde{r}_{B_{1,2}}^{(N)}$ that
  does go negative, thereby triggering the transition to a biaxial
  $N+2$ state.\cite{alban,deGennesProst}
\end{references}
\end{document}